\documentclass[reprint,superscriptaddress,english,twocolumn,prl]{revtex4-2}
\usepackage{xcolor}
\usepackage{babel}
\usepackage{amsmath}
\usepackage{amssymb}
\usepackage{graphicx}
\usepackage{lipsum}  
\usepackage{mathtools}
\usepackage{comment}
\usepackage{bbold}
\usepackage{braket}
\usepackage{hyperref}
\usepackage{url}
\usepackage{dsfont}
\usepackage{natbib}

\begin{document}

\preprint{APS/123-QED}

\title{Unifying Floquet theory of longitudinal and dispersive readout }

\author{A. Chessari}
\affiliation{IRIG-MEM-L\_Sim, Univ. Grenoble Alpes, CEA, Grenoble INP, Grenoble 38000, France}
\author{E. A. Rodr\'iguez-Mena}
\affiliation{IRIG-MEM-L\_Sim, Univ. Grenoble Alpes, CEA, Grenoble INP, Grenoble 38000, France}
\author{ J. C. Abadillo-Uriel} 
\affiliation{IRIG-MEM-L\_Sim, Univ. Grenoble Alpes, CEA, Grenoble INP, Grenoble 38000, France}
\affiliation{Instituto de Ciencia de Materiales de Madrid,
Consejo Superior de Investigaciones Cientificas,  Madrid 28049, Spain}
\author{V. Champain}
\affiliation{IRIG-Pheliqs, Univ. Grenoble Alpes, CEA, Grenoble INP, Grenoble 38000, France}
\author{S. Zihlmann}
\affiliation{IRIG-Pheliqs, Univ. Grenoble Alpes, CEA, Grenoble INP, Grenoble 38000, France}
\author{R. Maurand}
\affiliation{IRIG-Pheliqs, Univ. Grenoble Alpes, CEA, Grenoble INP, Grenoble 38000, France}
\author{Y.-M. Niquet}
\affiliation{IRIG-MEM-L\_Sim, Univ. Grenoble Alpes, CEA, Grenoble INP, Grenoble 38000, France}
\author{M. Filippone}
\affiliation{IRIG-MEM-L\_Sim, Univ. Grenoble Alpes, CEA, Grenoble INP, Grenoble 38000, France}

\begin{abstract}
We devise a Floquet theory of longitudinal and dispersive readout in circuit QED. By studying qubits coupled to cavity photons and driven at the resonance frequency of the cavity $\omega_{\rm r}$, we establish a universal connection between the qubit AC Stark shift and the longitudinal and dispersive coupling to photons. We find that the longitudinal coupling $g_\parallel$ is  controlled by the slope of the AC Stark shift as function of the driving strength $A_{\rm q}$, while the dispersive shift $\chi$ depends on its curvature. The two quantities become proportional to each other in the weak drive limit ($A_{\rm q}\rightarrow 0$). Our approach unifies the adiabatic limit ($\omega_{\rm r}\rightarrow 0$) -- where $g_\parallel$ is generated by the static spectrum curvature (or quantum capacitance) -- with the diabatic one, where the static spectrum plays no role.  We derive analytical results supported by exact numerical simulations. We apply them to superconducting and spin-hybrid cQED systems, showcasing the flexibility of faster-than-dispersive longitudinal readout. 
\end{abstract}

\maketitle

{\it Introduction -- }Improving and speeding-up readout is a key asset for the development of quantum architectures with multiple qubits. In the context of circuit quantum electrodynamics (cQED)~\cite{blais_cQED_2021}, fast 
measurements of the $\sigma_z$ state of a qubit rely on  dispersive readout~\cite{blais_cQED_2004}:  the lowest  two levels of a superconducting artificial atom exchange virtual excitations  with a detuned microwave resonator, enabling quantum non-demolition (QND) measurements.   
Speed and fidelity of dispersive readout increase with the power applied to the resonator, but are limited by leaking to high excited states and Purcell effect~\cite{walter_rapid_2017,shillito_dynamics_2022,cohen2023reminescence}. 

Recently, a longitudinal interaction of the form
~\cite{kerman2013quantum,billangeon2015cQED}
\begin{equation}\label{eq:longitudinal}
    \mathcal H_\parallel=g_\parallel(a+a^\dagger)\sigma_z
\end{equation} 
has emerged as an alternative solution for faster QND readout~\cite{didier_fast_2015,richer2016circuit,richer2017inductively,eichler2018realizing}\, where $a^{(\dagger)}$ annihilates (creates) photons in the resonator and $g_\parallel$ is the {\it longitudinal coupling}.  
Readout schemes based on Eq.~\eqref{eq:longitudinal} were implemented in cQED~\cite{touzard2019gated,ikonen2019qubit} and 
longitudinal coupling has been  probed in hybrid systems~\cite{clerk_2020_hybrid} featuring
spin~\cite{bottcher2022parametric} and charge qubits~\cite{corrigan2023longitudinal}. Recent demonstrations of strong spin-photon coupling~\cite{frey20212dipole,petersson2012circuit,viennot2015coherent,scarlino2022insitu,yu2023strong,depalma2023strong,kang2023coupling,dijkema2023twoqubit} pave the way to longitudinal readout of pure two-level qubits in spin cQED~\cite{harvey2018coupling,ruskov2019quantum,ruskov2021modulated,bosco2022fully,michal2023tunable,ruskov2023longitudinal}. 

\begin{figure}[b]
    \centering
    \includegraphics[width=\columnwidth]{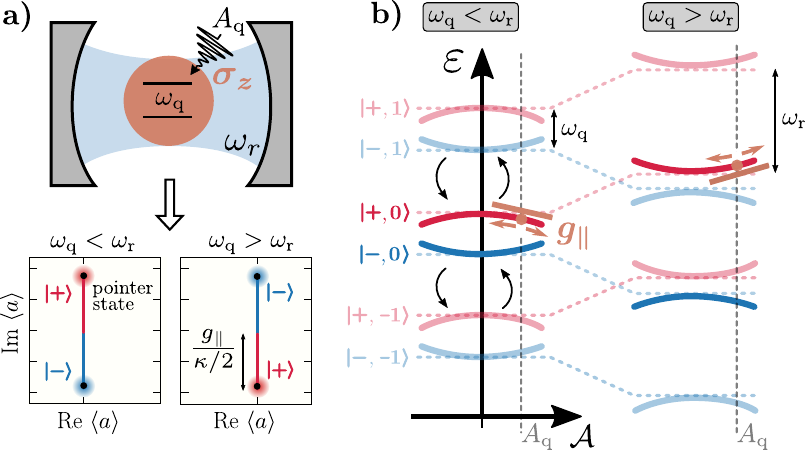}
    \caption{
    Longitudinal readout of qubits. a) The qubit is driven at the resonator frequency $\omega_{\rm r}$, which experiences an effective $\sigma_z$-dependent drive. In the case of pure longitudinal coupling~\eqref{eq:longitudinal}, the cavity pointer states move in opposite direction depending on the qubit state $|\pm \rangle$. b) The transverse drive to the qubit couples different replicas (labeled by integers) of the Floquet quasi-energy spectrum, leading to an AC Stark shift. The emergent longitudinal qubit-photon coupling $g_\parallel$ is given by the slope of the AC Stark shift at the driving strength $A_{\rm q}$. The sign of $g_\parallel$ depends on whether $\omega_{\rm r}\gtrless \omega_{\rm q}$.
    }
    \label{fig:setup}
\end{figure}

An identified mechanism to parametrically control the interaction~\eqref{eq:longitudinal} is sketched in Fig.~\ref{fig:setup}. Driving a qubit of Larmor frequency $\omega_{\rm q}$ at the resonator frequency $\omega_{\rm r}$ triggers a $\sigma_z$-dependent displacement of the resonator. However, it remains unclear which circuit properties  control the longitudinal coupling strength $g_\parallel$. For superconducting circuits,  $g_\parallel$ was found to connect to the dispersive shift $\chi$~\cite{ikonen2019qubit}, while,
in hybrid systems, a focus on adiabatic regimes ($\omega_{\rm r}\ll\omega_{\rm q}$) has established a connection between $g_\parallel$ and the {\it static curvature} of the Larmor frequency induced by variations of the drive amplitude $\partial^2\omega_q/\partial A_{\rm q}^2$ at equilibrium~\cite{harvey2018coupling,ruskov2019quantum,ruskov2021modulated,bosco2022fully,michal2023tunable,ruskov2023longitudinal}. As recently emphasized for dispersive readout~\cite{kohler2017dispersive,kohler2018dispersive,park_from_adiabatic_2020}, such situation calls for a unified description of cQED systems, bringing together different configurations and regimes, unrestricted by adiabatic or rotating-wave approximations.

In this work, we devise a unifying Floquet formalism to establish a universal connection between AC Stark shift, longitudinal and dispersive readout in cQED, see Fig.~\ref{fig:setup}. We show that the longitudinal coupling $g_\parallel$ is a generic mechanism, which is controlled by the {\it slope}, rather than the curvature, of the 
AC Stark shift caused by the drive strength $A_{\rm q}$. The curvature gives instead the dispersive shift $\chi$, and  $g_\parallel\propto\chi$ only for $A_{\rm q}\rightarrow0$. 
We thus establish the possibility of parametric longitudinal readout at any finite detuning $\delta=\omega_{\rm q}-\omega_{\rm r}$, discussing its advantages and limitations compared to dispersive readout, equally described by our approach, even at strong drives. 
We derive analytical results supported by exact numerics
, which we apply to superconducting and spin cQED settings, providing a unifying, seamless and simple description of 
generic readout mechanisms in 
cQED.


{\it Derivation --} To establish the relation between AC Stark shift,  longitudinal and dispersive readout, we start by considering the simplest  case of a (charge) qubit of Larmor frequency $\omega_{\rm q}$ transversally coupled to a resonator of frequency $\omega_{\rm r}$, and transversally driven at the same frequency, see Fig.~\ref{fig:setup}. This scenario is described by the time-periodic Rabi Hamiltonian (we assume $\hbar=1$) 
\begin{equation}\label{eq:ham}
\mathcal H=\frac{\omega_{\rm q}}2\,\sigma_z+\omega_{\rm r}\,a^\dagger a+g_\perp(a+a^\dagger)\sigma_x+A_{\rm q}\cos(\omega_{\rm r}t)\sigma_x\,,
\end{equation}
where the drive amplitude $A_{\rm q}$ acts as a static detuning in the adiabatic $\omega_{\rm r}\rightarrow0$ limit. In the frame rotating at frequency $\omega_{\rm r}$, the cavity field operators $a^{(\dagger)}$ renormalize the drive amplitude $A_{\rm q}$ according to the  Hamiltonian 
\begin{align}\label{eq:starting}
\mathcal H'&=\frac{\omega_{\rm q}}{2}\sigma_{z}+\frac12\left(\mathcal Ae^{-i\omega_{\rm r} t}+ \mathcal{A}^\dagger e^{i\omega_{\rm r} t}\right) \sigma_x\,, 
\end{align}
with $\mathcal A=A_{\rm q}+2g_\perp a$. For simplicity, we initially assume that, when the resonator loses photons at  rate $\kappa$, the driven qubit steers the cavity towards a coherent state $|\alpha\rangle$, such that $a|\alpha\rangle=\alpha|\alpha\rangle$. In this case, we can treat $\mathcal A$ as a number and simply derive the effective qubit-photon coupling within the Floquet formalism~\cite{shirley_floquet,sambe1973steady,GRIFONI_driven_1998,CHU_beyond_2004,rudner2020floquet}, which we review in the Supplemental Material (SM)~\cite{SM}.  

In analogy to space-periodic systems, Floquet's theorem states that the solutions of the Schr\"odinger equation corresponding to the time-periodic Hamiltonian~\eqref{eq:starting} read $\ket{\Psi_j(t)}=e^{-i\varepsilon_jt}\ket{u_j(t)}$, with periodic $\ket{u_j(t+2\pi/\omega_{\rm r})}=\ket{u_j(t)}$. The quasi-energies $\varepsilon_j$ are defined  up to multiple integers of $\omega_{\rm r}$, since $\ket{ u_j(t)}$ and $e^{in\omega_{\rm r}t}\ket{u_j(t)}$ are equivalent solutions for any integer $n$. Thus, it is natural to build an extended space of replicas of the $\ket{\pm}$ states, each one associated to a replica of the quasi-energy. For $\mathcal A=0$, the Floquet spectrum consists of the replicated qubit energies $\pm\omega_{\rm q}/2+n\omega_{\rm r}$, see Fig.~\ref{fig:setup}b. For $\mathcal A\neq0$, the time-periodic term in Eq.~\eqref{eq:starting} couples every $\ket\pm$ state of one replica to the $\ket\mp$ state of the closest one~\cite{SHEVCHENKO2010laundauzener,grifoni_dissipative_2010,deng2015observation,deng2016dynamics,koski2018floquet}.
Similarly to a charge qubit detuning,  $|\mathcal A|$ controls the repulsion between replicas, leading to the AC Stark shift of the qubit states. Thus, the quasi-energies acquire {\it curvature} as function of  $|\mathcal A|$. As sketched in Fig.~\ref{fig:setup}b, the sign of the detuning $\delta$ between qubit and resonator controls the sign of this curvature. The Floquet Hamiltonian $\mathcal H_{\mathcal F}$ associated to Eq.~\eqref{eq:starting} is the time-independent Hamiltonian whose eigenvalues are the quasi-energies $\varepsilon_j$. It can be cast in the diagonal form
\begin{equation}\label{eq:floquet}
\mathcal H_{\mathcal F} =\varepsilon\big(\omega_{\rm q},\omega_{\rm r},|\mathcal A|\big)\,\tilde\sigma_z\,,
\end{equation}
where $\tilde\sigma_z$ is the $z$ Pauli matrix,  acting on replica  $n=0$. 

For sufficiently weak occupations of the cavity ($g_\perp|\alpha|\ll A_{\rm q}$), we can expand the quasi-energy $\varepsilon$  close to $\mathcal A=A_{\rm q}$, up to second order in $g_\perp \alpha$. Neglecting all Kerr terms ($\propto \alpha^2,\alpha^{*2}$), we find
\begin{equation}\label{eq:effective_H}
\mathcal H_{\mathcal F} =\left[ \varepsilon(|A_{\rm q}|)+g_\parallel(  \alpha+ \alpha^*)+\chi  |\alpha|^2\right]\tilde\sigma_z\,,
\end{equation}
where
\begin{align}\label{eq:gperpchi}
\frac{g_\parallel}{g_\perp} &=  \left. \frac{\partial \varepsilon}{\partial  \mbox{Re} \mathcal A } \right\vert_{A_{\rm q}},&
\frac{\chi}{g^2_\perp} &= \left[ \frac{\partial^2 \varepsilon}{\partial |\mathcal A |^2} + \frac{1}{|\mathcal A|} \frac{\partial \varepsilon}{\partial | \mathcal A |} \right]_{A_{\rm q}}\,.\end{align}
These equations are the main result of our work. They establish a relation between the AC Stark shift, 
controlled by the driving amplitude $A_{\rm q}$, the longitudinal coupling $g_\parallel$ and the dispersive shift $\chi$. The longitudinal coupling $g_\parallel$ is proportional to the {\it slope} of the Floquet quasi-energies, while the dispersive coupling $\chi$  includes a contribution from their {\it curvature} with respect to the drive strength. 

Equation \eqref{eq:gperpchi} generalizes the expectation that both $g_\parallel$ and $\chi$ are governed by the curvature of the static spectrum $\partial^2 \varepsilon/\partial |\mathcal A|^2|_{A_{\rm q},\omega_{\rm r}\rightarrow0}$~\cite{didier_fast_2015,richer2016circuit,harvey2018coupling,ruskov2019quantum,ruskov2021modulated,bosco2022fully,michal2023tunable,ruskov2023longitudinal} or, equivalently, the quantum capacitance~\cite{park_from_adiabatic_2020}. Indeed, the static regime is directly obtained by considering Eq.~\eqref{eq:gperpchi} in the $\omega_{\rm r}\rightarrow0$ limit, where the drive in Eq.~\eqref{eq:ham} acts as a static detuning. 

If we instead consider the small drive limit $A_{\rm q}\rightarrow 0$, but for arbitrary $\omega_{\rm r}$, we  find a direct connection  between the longitudinal and the dispersive coupling
\begin{align}\label{eq:gpchi_small_Aq}
g_{\parallel}^{(0)}&= \frac{\chi^{(0)}}{2g_\perp}A_{\rm q}\,, &\chi^{(0)}&=2g_\perp^2\left.\frac{\partial^2\varepsilon}{\partial|\mathcal A|^2}\right|_{A_{\rm q }=0}\,. 
\end{align}
In this limit, both couplings become proportional to the Floquet (AC Stark) curvature $\partial^2\varepsilon/\partial|\mathcal A|^2|_{A_{\rm q }=0}$, and, in the specific case of a charge qubit, $\chi^{(0)}= 2g_\perp^2\omega_{\rm q }/(\omega_{\rm q}^2-\omega_{\rm r}^2)$. Equation~\eqref{eq:gpchi_small_Aq} showcases the parametric dependence of the longitudinal coupling $g_\parallel$ on the drive $A_{\rm q}$. The relation between $g^{(0)}_\parallel$ and $\chi^{(0)}$ was first found for transmons~\cite{ikonen2019qubit}. We show here that this relation  holds in general and that it is connected to Floquet spectral properties. Notice that both $g^{(0)}_\parallel$ and $\chi^{(0)}$  are proportional to
$\omega_q^{-1}$ in the adiabatic limit ($\omega_{\rm r}\rightarrow0$)~\cite{corrigan2023longitudinal,ruskov2023longitudinal}, corresponding to the static curvature of a charge qubit at zero detuning~\cite{park_from_adiabatic_2020,ruskov2023longitudinal} (see also SM~\cite{SM}), but change sign when $\omega_{\rm r}>\omega_{\rm q}$.

In the SM~\cite{SM}, we derive Eqs.~(\ref{eq:effective_H}-\ref{eq:gperpchi}) in the  operator formalism. We apply a Schrieffer-Wolff transformation~\cite{schrieffer1966relation} in Floquet space and derive the Hamiltonian~\eqref{eq:effective_H}, with the correct dispersive correction to the Larmor frequency
\begin{equation}\label{eq:effective_H_a}
\mathcal H_{\mathcal F}' =\left[ \varepsilon(|A_{\rm q}|)+\frac\chi2+g_\parallel(  a+ a^\dagger)+\chi  a^\dagger  a \right]\tilde\sigma_z\,.
\end{equation}

\begin{figure*}[!t]
    \centering
    \includegraphics[width=\textwidth]{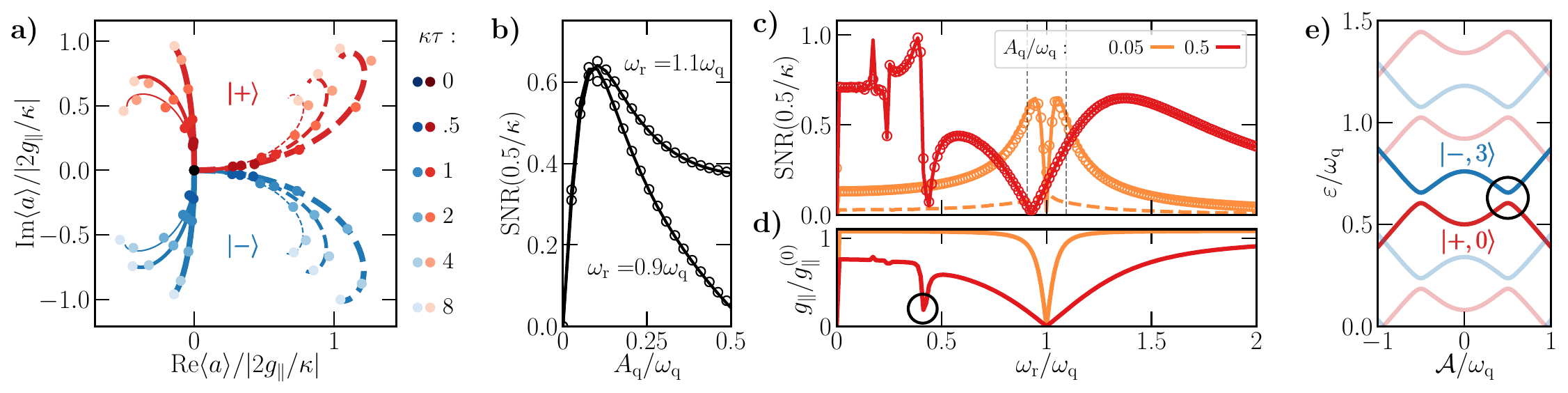}
    \caption{ a) Analytical time trajectories of $\langle a\rangle$ from the effective Hamiltonian~\eqref{eq:effective_H_a} for longitudinal (solid) and  dispersive readout (dashed). The dots correspond to numerical simulations of Eq.~\eqref{eq:ham}. Longitudinal trajectories are obtained for $A_{\rm q}/\omega_{\rm q}=0.05$ and $\omega_{\rm r}/\omega_{\rm q}=1.1,1.15,1.5$  (thin to thick). For dispersive readout, we set $\omega_{\rm r}/\omega_{\rm q}=1.1$, to enforce the condition $\vert \chi^{(0)} \vert=\kappa/2$, with $g_\perp /\omega_{\rm q}=10^{-2}$ and $\kappa/\omega_{\rm q}=2\cdot 10^{-3}$. The cavity drive $A_{\rm r}$ is adjusted accordingly to match the longitudinal SNR at $t\rightarrow\infty$. Dispersive trajectories, for which $g_\parallel=0$, are expressed in the same unit as the longitudinal counterpart. 
    b) Parametric increase and suppression of the SNR at time $t=0.5/\kappa$ as function of $A_{\rm q}$ for $\omega_{\rm r}/\omega_{\rm q}=0.9,1.1$ (vertical dashed lines in panel (c)). c) SNR  for different values of $\omega_{\rm r}/\omega_{\rm q}$ and drive strength $A_{\rm q}$. The dashed  line corresponds to the optimal dispersive case matching the longitudinal SNR in panel (a) for $A_{\rm q}/\omega_{\rm q}=0.05$. 
    d) Deviations from unity of $g_\parallel/g_\parallel^{(0)}$ as function of $\omega_{\rm r}/\omega_{\rm q}$. 
    e) Floquet spectrum as function of the renormalized drive $\mathcal A$ (assumed to be real) for $\omega_{\rm r }/\omega_{\rm q}=0.42$. The circle highlights avoided crossings between distant replicas at $A_{\rm q}\simeq0.5$, leading to the suppression of $g_\parallel$ in panel (d).
    }
    \label{fig:comparison}
\end{figure*}

{\it Validity and comparison with dispersive readout -- }We benchmark our approach on exact numerical simulations in Fig.~\ref{fig:comparison}~\footnote{The numerical calculations were performed with QuTiP~\cite{JOHANSSON_qutip_2013} and scQubits~\cite{scqubits1,scqubits2} for the superconducting devices, with details provided in the SM~\cite{SM}. All the codes for generating data can be found at \href{https://github.com/aleeschessari/Unifying-Floquet-Theory}{https://github.com/aleeschessari/Unifying-Floquet-Theory}}. There, we compare the $\sigma_z$-dependent dynamics of the cavity field $\langle a\rangle$ according to  longitudinal or standard dispersive readout protocols, where either the qubit or the cavity are driven at frequency $\omega_{\rm r}$. Dispersive readout is also described by Eq.~\eqref{eq:effective_H_a} with $A_{\rm q}=\,g_\parallel=0$. The driving term is instead $A_{\rm r}\sin(\omega_{\rm r}t+\pi)(a+a^\dagger)$~\footnote{We use the sine instead of the cosine in Eq.~\eqref{eq:ham}
to allow a clearer comparison with longitudinal coupling in Fig.~\ref{fig:comparison}.}, which populates the cavity to probe its $\sigma_z$-dependent frequency shift  $\omega_{\rm r}'=\omega_{\rm r}+\langle\sigma_z\rangle\chi/2$~\cite{blais_cQED_2004,blais_cQED_2021}, see SM~\cite{SM}. 

Figure~\ref{fig:comparison}a compares 
longitudinal and dispersive readout 
for $\omega_{\rm r}>\omega_{\rm q}$. The parameters for both protocols have been chosen such that the signal-to-noise ratios, 
$\text{SNR}^2(t)=2\kappa\int_0^td\tau | \langle  a(\tau) \rangle_{+}-\langle  a(\tau) \rangle_{-} |^2$, coincide in the steady state ($t\rightarrow\infty$)~\cite{didier_fast_2015}. We focus on the SNR, which provides a unique, and experimentally relevant~\cite{blais_cQED_2021,didier_fast_2015}, figure of merit for readout quality over different detunings and driving strengths.
We derive analytically the time evolution of $\langle a\rangle$ from the effective model~\eqref{eq:effective_H_a}. The stationary pointer states fulfill approximately $\langle a\rangle_{\sigma_z}\simeq(-ig_\parallel\langle\sigma_z\rangle+A_{\rm r}/2)/(i\chi\langle\sigma_z\rangle+\kappa/2)$~\cite{didier_fast_2015,corrigan2023longitudinal}\footnote{This expression neglects entanglement between qubit and cavity, which has to be taken into account to get a good agreement with exact numerics at strong drives. The general expression is given in the SM~\cite{SM}.}. Notice that 
in the opposite regime $\omega_{\rm r}<\omega_{\rm q}$, the trajectories associated to $\ket\pm$ in Fig.~\ref{fig:comparison}a  simply switch,  according to the predicted sign change of $g_\parallel$ and $\chi$, see SM~\cite{SM}.  

The longitudinal dynamics is clearly affected by the dispersive coupling $\chi$, but it is entirely different from the dispersive one. At short times,  the pointer states of the cavity evolve as $\langle  a(t) \rangle_{\rm long.} \approx -ig_\parallel \langle \sigma_z(0) \rangle t$. They thus split immediately at $180^\circ$ while, in the dispersive case, the short-time trajectories of the resonator are  
$\sigma_z$-independent: $\langle  a(t) \rangle_{\rm disp.} \approx A_{\rm r}t$. This difference makes longitudinal readout faster than dispersive~\cite{didier_fast_2015}.

For dispersive readout, we always consider the condition $|\chi|=\kappa/2$, where pointer states maximally split at all times~\cite{blais_cQED_2004,blais_cQED_2021}. However, this condition has no special meaning for longitudinal readout. Figure~\ref{fig:comparison}a shows that purest longitudinal splitting is obtained by suppressing $\chi$, which is achieved increasing the qubit-cavity detuning, see Eq.~\eqref{eq:gpchi_small_Aq}. This expression also predicts that $g_\parallel$ vanishes with $\chi$. However, this suppression  can be compensated by increasing $A_{\rm q}$.

Figure ~\ref{fig:comparison}a shows the good agreement between the Floquet formalism and the exact numerics. 
The quality of the agreement also appears
in Fig.~\ref{fig:comparison}b-c, which shows the overall dependence of the SNR at $t=0.5/\kappa$ on the drive amplitude $A_{\rm q}$ and on the ratio $\omega_{r}/\omega_{\rm q}$. As expected, longitudinal readout outperforms the dispersive one at short times (roughly a factor 5 at  $t=0.5/\kappa$)~\cite{didier_fast_2015}. 

Beyond the real-time dynamics, the Floquet approach accurately and simply describes complex processes emerging for large driving strengths $A_{\rm q}$. For instance, Figs.~\ref{fig:comparison}b-c show that the parametric increase of the SNR  stops at large drives, with sudden drops for specific  values of $\omega_{\rm r}$. These drops are entirely mirrored by the corresponding suppression of $g_\parallel$ in Fig.~\ref{fig:comparison}d. The inspection of the Floquet spectrum  in Fig.~\ref{fig:comparison}e shows that this suppression is caused by resonant Landau-Zener-St\"uckelberg transitions between the $\ket{\pm}$ qubit states~\cite{SHEVCHENKO2010laundauzener}. They are signaled by anti-crossings between distant replicas, which flatten the Floquet spectrum and, according to Eq.~\eqref{eq:gperpchi}, suppress $g_\parallel$. Such resonances appear at much larger values of $A_{\rm q}$ for $\omega_{\rm r}>\omega_{\rm q}$, see SM~\cite{SM}. We emphasize that operating in this regime, where the notion of static curvature~\cite{harvey2018coupling,bottcher2022parametric,corrigan2023longitudinal,ruskov2019quantum,ruskov2021modulated,bosco2022fully,michal2023tunable,ruskov2023longitudinal} cannot be applied, has also the  advantage to suppress Purcell relaxation caused by spurious higher harmonics of the resonator.

We conclude stressing that Fig.~\ref{fig:comparison}d shows the limited range of validity of the $A_{\rm q}\rightarrow0$ estimation of $g_{\parallel}^{(0)}$, Eq.~\eqref{eq:gpchi_small_Aq}. The deviations of $g_\parallel/g^{(0)}_\parallel$ from one show that $g^{(0)}_\parallel$ greatly overestimates  $g_\parallel$, showcasing the effectiveness of the Floquet approach for correct estimations.

Our analysis is relevant for actual devices, as these results directly apply to the readout of charge qubits at zero detuning~\cite{corrigan2023longitudinal}, which however suffer from short coherence times~\cite{nakamura1999coherent,hayashi2003coherent,shinkai2009correlated,petersson2010quantum,dovzhenko2011nonadiabatic,shi2013coherent,cao2013ultrafast,kim2015microwave,scarlino2022insitu}. We illustrate below that our approach readily extends to realistic spin and superconducting cQED systems.

{\it Extension to multi-level systems -- 
} 
Realistic systems require to go beyond the idealized two-level scenario. 
In the SM~\cite{SM}, we show that a straightforward generalization 
of Eqs.~(\ref{eq:floquet}-\ref{eq:effective_H_a}) to $m$-levels still holds~\cite{ruskov2023longitudinal}. The effective Floquet Hamiltonian becomes  
\begin{equation}\label{eq:effective_H_a_multi}
\mathcal H_{\mathcal F}''=\sum_{j=1}^m\left[ \varepsilon_j(|A_{\rm q}|)+\frac{\chi_j}{2}+g_{\parallel,j}(  a+ a^\dagger)+\chi_j a^\dagger  a \right]\tilde{\mathcal P}_j\,,
\end{equation}
where $\tilde{\mathcal P}_j$ projects onto the $j$-th  Floquet eigenmode. Each mode couples independently to the cavity with its own longitudinal $g_{\parallel, j}$ and dispersive $\chi_j$. They are both given by Eq.~\eqref{eq:gperpchi} replacing 
$\varepsilon\rightarrow\varepsilon_j$. In general $\varepsilon_1(|\mathcal A|)\neq-\varepsilon_0(|\mathcal A|)$. Thus, projecting on the two level subspace $\ket\pm$,  the mapping onto Eq.~\eqref{eq:effective_H_a_multi} leads to an effective longitudinal coupling $ g_\parallel=\left(g_{\parallel,1}-g_{\parallel,0}\right)/2$ plus an additional $\sigma_z$-independent term $\bar g_\parallel=\left(g_{\parallel,1}+g_{\parallel,0}\right)/2$. Its effects  are discussed in the applications below. 


\begin{figure}
    \centering
    \includegraphics[width=\columnwidth]{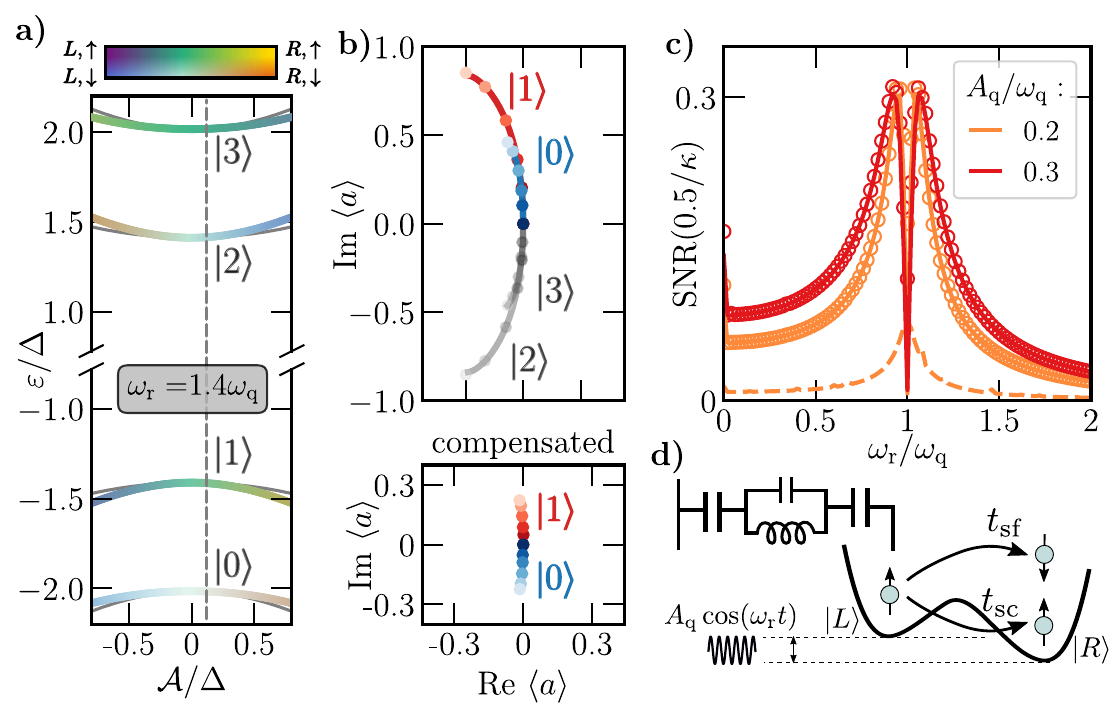}
    \caption{Longitudinal readout of spin qubits. a) Floquet spectra as function of the renormalized drive $\mathcal A$ (assumed to be real). The qubit frequency $\omega_{\rm q}$ is set by the difference between the $\ket0$ and $\ket1$ qubit states for $A_{\rm q}=0$. The grey lines show the static limit $\omega_{\rm r}=0$, where $A_{\rm q}$ acts as a static detuning. b) Longitudinal splitting of the resonator pointer states  depending on the initial state of the system for $\omega_{\rm r}/\omega_{\rm q}=1.4$ and $A_{\rm q}/\omega_{\rm q}=0.2$ (value corresponding to the dashed vertical lines in panel (a)). The dynamics given by $\bar{g}_\parallel$ can be compensated by an additional drive on the cavity (bottom panel). Dots correspond to numerical simulations, as in Fig.~\ref{fig:comparison}a. c) SNR as in Fig.~\ref{fig:comparison}c.  
    d) Experimental setup.
    Model parameters (in units of $\Delta$): $t_{\text{sc}}=1$,  $t_{\text{sf}}=1.3$, $g_\perp=2\cdot 10^{-2}$, $\kappa=2\cdot10^{-3}$, $\omega_{\rm q}=0.6$~\cite{yu2023strong}.
    }
    \label{fig:flopping_double_case}
\end{figure}

{\it  Hybrid spin cQED --}  
We consider a double quantum dot trapping a single charge and coupled to a superconducting  cavity, see  Fig.~\ref{fig:flopping_double_case}. Spin-orbit (SO) interactions convert the strong charge-photon coupling to spin-photon coupling~\cite{Jin2012,frey20212dipole,2012_Petersson,viennot2015coherent,2018_Samkharadze,2018_Mi,2019_Borjans,2020_Harvey-Collard,yu2023strong,beaudoin2016coupling}. We detail in the SM~\cite{SM} the well-known models describing such devices~\cite{2004_Childress,Burkard2006,trif2008spin,cottet2010spin,2012_Hu}, choosing parameters from actual experimental conditions~\cite{yu2023strong}.  

For the present discussion, we are interested in the structure of the Floquet spectrum of the system in Fig.~\ref{fig:flopping_double_case}a. It is composed of four branches, mixing left and right $\ket{L,R}$ and spin $\ket{\uparrow,\downarrow}$ degrees of freedom. The Zeeman energy $\Delta$ splits the $\ket{\uparrow,\downarrow}$ states, and the amplitudes $t_{\rm sc/sf}$ describe spin-conserving/flipping tunneling between the dots. SO is necessary for finite $t_{\rm sf}$. 

We focus here on 
flopping-mode regimes~\cite{mutter_natural_2021}, where the Zeeman splitting  is dominated by 
spin-conserving tunneling ($\Delta<2t_{\rm sc}$).  Figure~\ref{fig:flopping_double_case}a shows that, differently from  charge qubits, the Floquet quasi-energies of the $\ket{0,1}$ logical states  have slopes and curvatures of the same sign. 
Thus, the multi-level generalization~\eqref{eq:effective_H_a_multi} implies  that the pointer states 
move in the same direction controlled by the average $\bar g_{\parallel}$, but still split longitudinally according to the slope difference $g_\parallel$, see Fig.~\ref{fig:flopping_double_case}b (top panel). However, the application of a compensation tone to the resonator 
eliminates the $\bar g_\parallel$ term~\cite{ikonen2019qubit} and restores longitudinal dynamics, see Fig.~\ref{fig:flopping_double_case}b (bottom panel).  
We illustrate below that the situation is similar for transmons and fluxonia.


\begin{figure}[t!]
    \centering
    \includegraphics[width=\columnwidth]{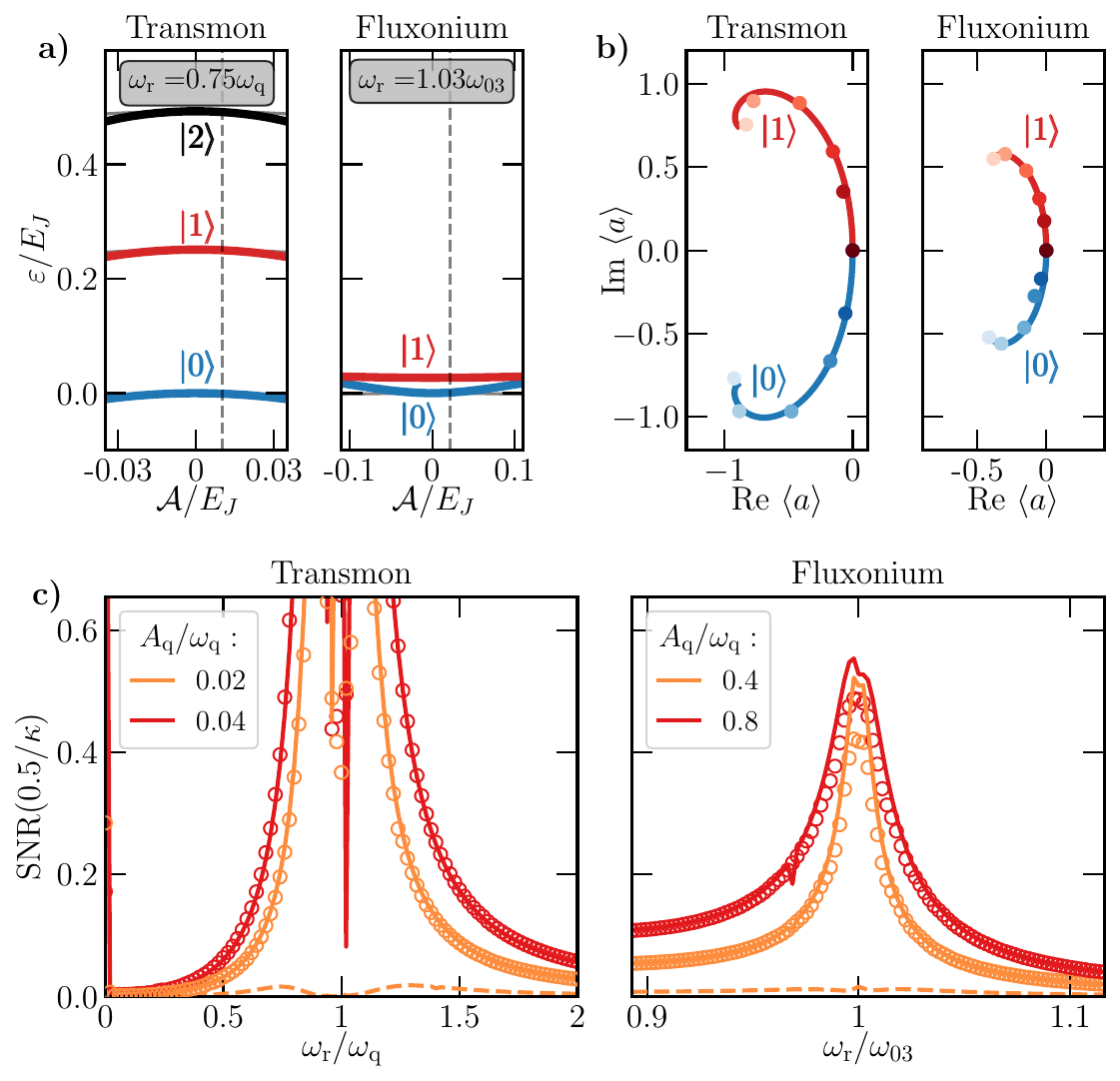}
    \caption{Longitudinal readout of transmon and fluxonium qubits. a) Floquet spectra ($\mathcal A$ assumed to be real) compared to the static limit at $\omega_{\rm r}=0$ (gray thin lines).  b) Evolution of the pointer states of the cavity in the presence of a compensation tone (analytical lines and numerical dots). The transmon is driven with  $A_{\rm q}/\omega_{\rm q}=0.04$ and the fluxonium with $A_{\rm q}/\omega_{\rm q}=0.8$ (vertical dashed lines in panel (a)). c) SNR at $t=0.5/\kappa$ as in Fig.~\ref{fig:comparison}c. Model parameters (in units of the Josephson Energy $E_J$): $g_\perp=3\cdot10^{-3}$, $\kappa=6.4\cdot 10^{-5}$, $E_C = 8.4\cdot 10^{-3}$, $\omega_{\rm q}=0.25$ for the transmon ($E_J/h=31.3$ GHz)~\cite{ikonen2019qubit} and $g_\perp=1\cdot10^{-2}$, $\kappa=1\cdot10^{-3}$, $E_C = 0.2$, $E_L=0.1$, $\Phi_{\rm ext}/\Phi_0 = 0.5$, $\omega_{\rm q}=0.03$ for the fluxonium ($E_J/h=5.57$ GHz)~\cite{somoroff2023millisecond}. Additional regimes for transmon and fluxonium are discussed in the SM~\cite{SM}. }
    \label{fig:fluxonium_transmon}
\end{figure}

\textit{Transmons and Fluxonia --} We conclude by applying 
our formalism 
to the longitudinal readout of superconducting transmon~\cite{koch2007charge-insensitive,schreier2008suppressing} and fluxonium~\cite{manucharyan2009fluxonium,somoroff2023millisecond} qubits driven transversally. Models and parameters describing 
such systems are extensively discussed in the literature and we provide all the details  in the SM~\cite{SM}.

Figure~\ref{fig:fluxonium_transmon}a shows that, similarly to the previous case, the Floquet quasi-energies of the $\ket{0,1}$ logical states have slopes and curvatures of the same sign. 
Also in this case,  the application of a compensation tone to the resonator eliminates the $\bar g_\parallel$ term~\cite{ikonen2019qubit} and restores longitudinal dynamics leading faster-than-dispersive readout, see Fig.~\ref{fig:fluxonium_transmon}b.  The weak anharmonicity of the transmon leads to additional resonances for $\omega_{\rm r}\sim\omega_{\rm q}$, spoiling readout when increasing the drive amplitude. 
These resonances are absent for the fluxonium, which we inspect for a resonator close to the $|0\rangle\rightarrow|3\rangle$ transition~\cite{somoroff2023millisecond}. 
The SNR in Fig.~\ref{fig:fluxonium_transmon}c shows a factor 5 gain at short times, 
compared to dispersive readout.

{\it Conclusions --}  In this work, we have devised 
a Floquet formalism  to connect 
the AC Stark shift of a driven qubit (or multi-level system) with 
its 
longitudinal and dispersive couplings to cavity photons. Our approach provides a simple and seamless  description of various regimes within a unifying framework,  
extending previous approaches~\cite{didier_fast_2015,touzard2019gated,corrigan2023longitudinal,ruskov2019quantum,ruskov2021modulated,bosco2022fully,michal2023tunable,ruskov2023longitudinal,park_from_adiabatic_2020,kohler2017dispersive,kohler2018dispersive}, and providing new  analytical results, supported by exact numerics, which straightforwardly generalize to various cQED platforms, ranging from hybrid to purely superconducting devices. 

Our approach brings new physical understanding by going beyond the notion of curvature coupling for $g_{\parallel}$~\cite{didier_fast_2015,ruskov2019quantum,ruskov2021modulated,bosco2022fully,michal2023tunable,ruskov2023longitudinal,park_from_adiabatic_2020,kohler2017dispersive,kohler2018dispersive}. Equipped with this understanding, we demonstrate longitudinal readout regimes where there is protection against resonator-mediated qubit relaxation. By showing the reduction of the SNR at large drives, we have also established a new connection between readout degradation and the flattening of the Floquet spectrum.


Even tough we have focused here on longitudinal readout induced by transverse driving, our approach can be in principle extended to account for polarizability effects~\cite{ruskov2023longitudinal}  and describe other protocols~\cite{dassoneville2020fast,lledo2023cloaking,munozarias2023qubit}, including the study of ZZ gates mediated by superconducting resonators~\cite{richer2016circuit}. Open questions concern how such Floquet methods can provide insight in strongly driven-regimes~\cite{shillito_dynamics_2022}, where photons are strongly entangled to the system and perturbative approaches fail.

\begin{acknowledgments}
{\it Acknowledgments --}  We are grateful to Quentin Ficheux, Benjamin Huard, Vincent Michal, Alexandru Petrescu,  Adria Rospars and Rusko Ruskov for helpful comments and discussions. We thank in particular Adria Rospars for bringing our attention to the fluxonium. A. C. and M. F. acknowledge support from EPiQ ANR-22-PETQ-0007 part of Plan France 2030C. JCAU is supported by a fellowship from the Fundaci\'on General
CSIC's ComFuturo programme which has received funding from the EU's Horizon 2020 research and innovation program under the Marie Skłodowska-Curie grant agreement No. 101034263.
\end{acknowledgments}

\bibliographystyle{apsrev4-2}
\bibliography{biblio}

\pagebreak
\clearpage

\onecolumngrid
\begin{center}
  \textbf{\large Supplemental Material to ``Unifying Floquet theory of longitudinal and dispersive  readout''}\\[.1cm]
  A. Chessari,$^1$ E. A. Rodr\'iguez-Mena,$^1$, J. C. Abadillo-Uriel,$^{1,2}$ V. Champain,$^3$ S. Zihlmann,$^3$ R. Maurand$^3$, Y.-M. Niquet$^1$, and Michele Filippone$^1$\\
  
  {\itshape $^1$IRIG-MEM-L\_Sim, Univ. Grenoble Alpes, CEA, Grenoble INP, Grenoble 38000, France\\
  $^2$Instituto de Ciencia de Materiales de Madrid,
Consejo Superior de Investigaciones Cientificas,  Madrid 28049, Spain\\
  $^3$IRIG-Pheliqs, Univ. Grenoble Alpes, CEA, Grenoble INP, Grenoble 38000, France\\
  }
(Dated: \today)\\[1cm]
\end{center}


\setcounter{equation}{0}
\setcounter{figure}{0}
\setcounter{table}{0}
\setcounter{page}{1}
\makeatletter
\renewcommand{\theequation}{S\arabic{equation}}
\renewcommand{\thefigure}{S\arabic{figure}}
\renewcommand{\bibnumfmt}[1]{[#1]}
\renewcommand{\citenumfont}[1]{#1}

In this Supplemental Material we give a short introduction on Floquet theory to show how it is applied for the derivation of Eqs. (4-7) in the main text. We then show how a Schrieffer-Wolff transformation allows to extend the validity of these results for non-coherent states of the cavity, leading to the Hamiltonian~(8) in the main text. We provide then details about the analytical derivation of the cavity state evolution based on this effective model and its benchmark on exact numerical simulations. We discuss the generalization of these results to multi-level systems and its application to spin-hybrid and superconducting cQED devices. 

\section{Derivation of Eqs.~(4-7) based on the Floquet formalism}
\label{sec:floquet_theory}
We provide here a short review of the Floquet formalism in relation to the solution of the time-dependent Hamiltonian~(3) in the main text and then derive Eqs.~(4-7).
\subsection{Floquet formalism}
Consider the time-periodic Hamiltonian~(3) in the main text 
\begin{equation}\label{eq:ham_sup}
\mathcal H'=\frac{\omega_{\rm q}}{2}\sigma_{z}+\frac12\left(\mathcal Ae^{-i\omega_{\rm r} t}+\mathcal A^* e^{i\omega_{\rm r} t}\right) \sigma_x\,,
\end{equation}
where we have already made the assumption that $\mathcal A$ is a complex number. Floquet's theorem~\cite{shirley_floquet,sambe1973steady,GRIFONI_driven_1998,CHU_beyond_2004,rudner2020floquet} states that the solutions of the periodic time-dependent Hamiltonian~\eqref{eq:ham_sup} have the form 
\begin{align}\label{eq:states}
\ket{\psi_j\left(t\right)}&=e^{-i\varepsilon_{j}t}\ket{u_{j}\left(t\right)}\,, & \ket{u_{j}(t)}=\ket{u_{j}(t+2\pi/\omega_{\rm r})}\,.
\end{align} 
We refer to the states $\ket{u_{j}(t)}$ as Floquet modes. In strict analogy to Bloch's states in space-periodic systems, the Floquet modes $\ket{u_{j}(t)}$ share the time-periodicity of the Hamiltonian~\eqref{eq:ham_sup}. The Schr\"odinger equation for the $j$-th Floquet mode reads
\begin{equation}
(\mathcal H^\prime-i\partial_t)\ket{u_{j}\left(t\right)}=\varepsilon_{j}\ket{u_{j}\left(t\right)}.
\label{eq:floquet_schro}
\end{equation}
In our case, each Floquet mode decomposes as $\ket{u_j(t)}=\sum_{\sigma}u_{j,\sigma}(t)\ket\sigma$, where $\sigma=\pm$ and the coefficients obey the periodicity condition $u_{j,\sigma}(t+2\pi/\omega_{\rm r})=u_{j,\sigma}(t)$.  Thus, each Floquet mode can be decomposed in Fourier harmonics $u_{j,\sigma}(t)=\sum_{p=-\infty}^{\infty}e^{-ip\omega_{\rm r} t}u_{j,\sigma,p}$. Introducing the compact notation  for the Fourier coefficients $\ket{u_{j,p}}=\sum_\sigma u_{j,\sigma,p}\ket\sigma$, Eq.~\eqref{eq:floquet_schro} can be recast as 
\begin{equation}\label{eq:floquet_0}
\left( \frac{\omega_{\rm q}}{2}\sigma_z - p\, \omega_{\rm r} \right)\ket{u_{j,p}}+ \frac{\mathcal A}{2}\sigma_x\ket{u_{j,p-1}} + \frac{\mathcal A^*}{2}\sigma_x\ket{u_{j,p+1}}=\varepsilon_j\ket{u_{j,p}}\,,
\end{equation}
which is a time-independent problem. To understand the structure of the solutions of Eq.~\eqref{eq:floquet_0}, it is important to stress that if  $\ket{u_j(t)}$, in Eq.~\eqref{eq:states}, solves the time-dependent Schr\"odinger equation with quasi-energy $\varepsilon_j$, then also $\ket{u_j^{(n)}(t)}=e^{in\omega t}\ket{u_j(t)}$ with $\varepsilon_j^{(n)}=\varepsilon_j+n\omega_{\rm r}$ does. This is immediately demonstrated by the  equality
\begin{equation}
\ket{\psi_j\left(t\right)}=e^{-i\varepsilon_{j}t}\ket{u_{j}\left(t\right)}=e^{-i\varepsilon^{(n)}_{j}t}\ket{u^{(n)}_{j}\left(t\right)}\,,
\end{equation}
which implies, by making the Fourier decomposition $\ket{u^{(n)}_{j}(t)}=\sum_pe^{-ip\omega_{\rm r}t}\ket{u^{(n)}_{j,p}}$, that 
\begin{equation}\label{eq:shift}
\ket{u^{(n)}_{j,p}}=\ket{u_{j,p+n}}\,.
\end{equation}
As a consequence, Eq.~\eqref{eq:floquet_0} admits an infinite amount of equivalent solutions, called {\it Floquet replicas}. They are shifted in Fourier space according to Eq.~\eqref{eq:shift} and are associated to evenly spaced quasi-energies $\varepsilon_j^{(n)}$. We label the replicas by the index $n$, and we will take the convention that the $n=0$ replica is the coinciding with $\varepsilon_\pm=\pm\omega_{\rm q}/2$ for $\mathcal A=0$. To account for the replicated structure~\eqref{eq:shift}, it is useful to introduce an extended Hilbert space stacking up all the Fourier coefficients of a given replica $n$
\begin{align}\label{eq:floquet_notation}
\ket{\left.u_{j},n\right\rangle }&\equiv
\begin{pmatrix} 
\ldots\\
\ket{u^{(n)}_{j,-1}}\\
\ket{u^{(n)}_{j,0}}\\
\ket{u^{(n)}_{j,+1}}\\
\ldots\\
\end{pmatrix}
=\sum_p|p)\otimes\ket{u^{(n)}_{j,-p}}=\sum_p|p)\otimes\ket{u^{(0)}_{j,n-p}}=\begin{pmatrix} 
\ldots\\
\ket{u^{(0)}_{j,n-1}}\\
\ket{u^{(0)}_{j,n}}\\
\ket{u^{(0)}_{j,n+1}}\\
\ldots\\
\end{pmatrix}\,, 
\end{align}
where we have introduced a canonical basis for the extended replica space
\begin{align}
\ldots && \left\vert -1 \right)&=
\begin{pmatrix}
\vdots\\
0\\
0\\
1\\
\vdots
\end{pmatrix}\,,&
\left\vert 0 \right)&=
\begin{pmatrix}
\vdots\\
0\\
1\\
0\\
\vdots
\end{pmatrix}\,,&
\left\vert 1 \right)&=
\begin{pmatrix}
\vdots\\
1\\
0\\
0\\
\vdots
\end{pmatrix}\,,&\ldots
\end{align}
To stress the consistency of this notation, notice that the vectors $|u_{j},n\rangle\rangle$ are orthonormal
\begin{equation}
\delta_{jk}\delta_{nm}=\langle\langle u_{j},n \vert u_{k},m\rangle\rangle =\sum_{p} \braket{u^{(0)}_{j,n+p}\vert u^{(0)}_{k,m+p}}=\sum_{p} \braket{u_{j,p}\vert u_{k,p+m-n}}\,,
\label{eq:orthonormality}
\end{equation}
from which we deduce that the modes $\ket{u_j(t)}$ are orthonormal at any instant $t$
\begin{equation}
\begin{aligned}
\braket{u_j(t) \vert u_k(t)}&=\sum_{pq}e^{i(p-q)\omega_{\rm r}t}\braket{u^{(0)}_{j,p}\vert u^{(0)}_{k,q}}=\sum_p\braket{u^{(0)}_{j,p}\vert u^{(0)}_{k,p}}+\sum_{d\neq 0} e^{id\omega_{\rm r}t}\sum_{p}\braket{u^{(0)}_{j,p}\vert u^{(0)}_{k,p+d}}=\delta_{jk}\,,
\end{aligned}
\end{equation}
as it should be. Thus, one can exploit the relation $(p|u_j,n\rangle\rangle=\ket{u_{j,-p}^{(n)}}$  to  cast Eq.~\eqref{eq:floquet_0} as an eigenvalue problem 
\begin{equation}
\mathcal{H}_{\mathcal F} \ket{\left.u_{j},n\right\rangle } = \varepsilon_{j}^{(n)}\ket{\left.u_{j},n\right\rangle }\,,
\label{eq:floquet_eigenvalue_problem}
\end{equation}
where we have defined the Floquet Hamiltonian
\begin{equation}\label{eq:Floquet_H}
\mathcal{H}_{\mathcal F}= \mathbf{p}\,\omega_{\rm r}+\frac{\omega_{\rm q}}{2}\sigma_z+\frac{1}{2}\left( \mathcal A\,f_-+\mathcal A^*\,f_{+} \right)\sigma_x\,,
\end{equation}
and introduced diagonal jump operators in the extended space
\begin{align}
\mathbf{p}&= \sum_p p \left\vert p \right)\left( p \right\vert\,, & f_{-}&= \sum_p \vert p-1 )(p\vert\,. & f_{+}&= \sum_p \vert p+1 )(p\vert\,.
\end{align}

\subsection{Floquet spectrum and derivation of Eqs.~(4-6) in the main text}
\label{sec:numerical_spectrum}
\begin{figure}[t]
    \centering
    \includegraphics[width=.6\textwidth]{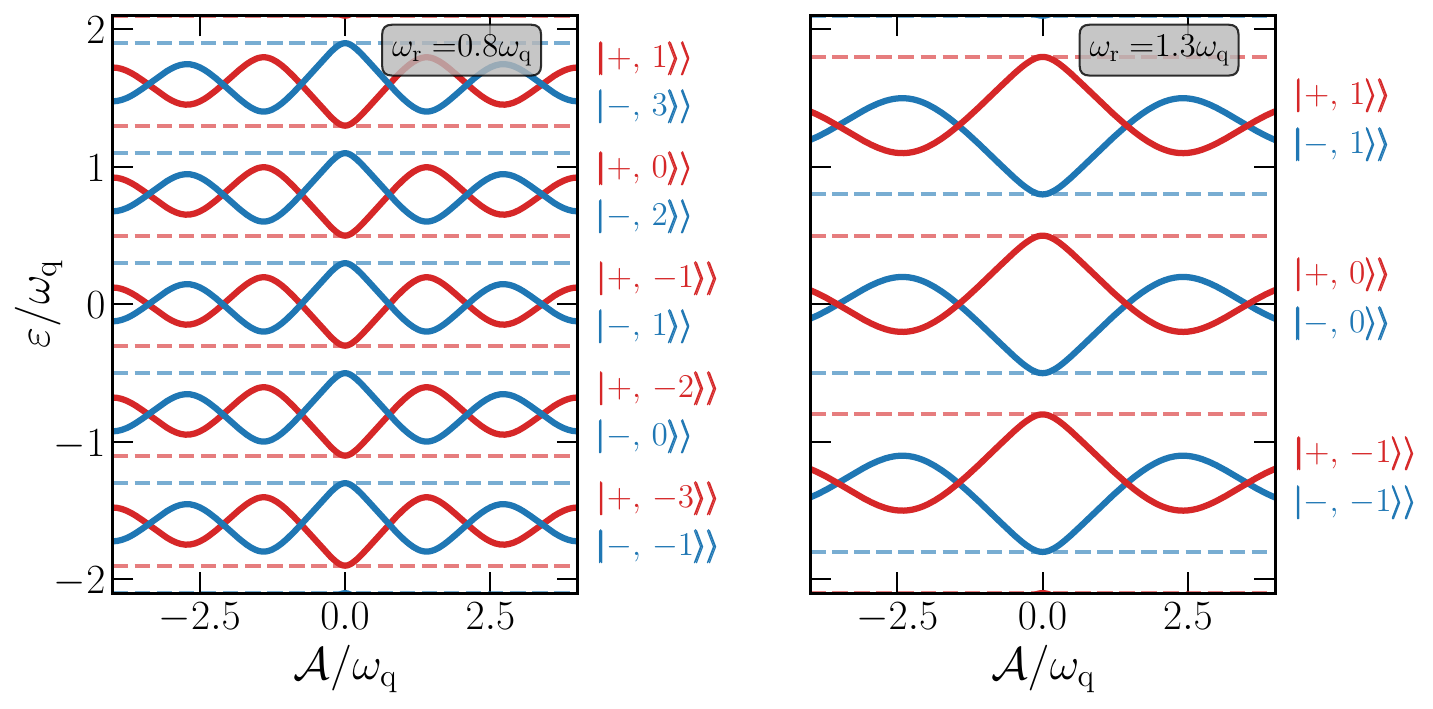}
    \caption{Floquet spectrum numerically computed considering $N_{\rm rep}=41$ replicas in the Floquet Hamiltonian~\eqref{eq:floquet_matrix} as function of the drive $\mathcal A$, that we chose to be real. We consider the case $\omega_{\rm r}<\omega_{\rm q}$ on the left and $\omega_{\rm r}>\omega_{\rm q}$ on the right, to highlight the change of curvature of the replicas. On the right side of each plot, we connect each quasi-energy $\varepsilon_j^{(n)}$ with its associated Floquet eigenstate $|u_j,n\rangle\rangle$. }
    \label{fig:floquet_charge_spectrum}
\end{figure}

The matrix representation of the Floquet Hamiltonian~\eqref{eq:Floquet_H} reads
\begin{equation}
\mathcal H_{\mathcal F}=
\begin{pmatrix}\ldots\\
 & \omega_{\rm r}+\frac{\omega_{\rm q}}{2} & 0 & 0 & \frac{\mathcal A^*}{2} & 0 & 0\\
 & 0 & \omega_{\rm r}-\frac{\omega_{\rm q}}{2} & \frac{\mathcal A^*}{2} & 0 & 0 & 0\\
 & 0 & \frac{\mathcal A}{2} & \frac{\omega_{\rm q}}{2} & 0 & 0 & \frac{\mathcal A^*}{2} \\
 & \frac{\mathcal A}{2} & 0 & 0 & -\frac{\omega_{\rm q}}{2} & \frac{\mathcal A^*}{2} & 0\\
 & 0 & 0 & 0 & \frac{\mathcal A}{2} & -\omega_{\rm r}+\frac{\omega_{\rm q}}{2} & 0\\
 & 0 & 0 & \frac{\mathcal A}{2} & 0 & 0 & -\omega_{\rm r}-\frac{\omega_{\rm q}}{2}\\
 &  &  &  &  &  &  & \ldots
\end{pmatrix}\,.
\label{eq:floquet_matrix}
\end{equation}
This matrix is diagonalized numerically by truncating the replica space up to  a number $N_{\rm rep}$ of replicas large enough to ensure convergence of the spectrum. Calculations of the quasi-energies $\varepsilon^{(n)}_{j}$ are shown in Fig.~\ref{fig:floquet_charge_spectrum} as function of $\mathcal A$ for $\omega_{\rm r}\gtrless\omega_{\rm q}$. These calculations show the change of the curvature of the spectrum sketched in Fig.~1 in the main text.

For the particular case of the charge qubit described by the Hamiltonian~\eqref{eq:ham_sup}, the diagonalized Floquet Hamiltonian takes the form
\begin{equation}\label{eq:diag1}
\mathcal H_{\mathcal F}=\tilde{\mathbf{n}}\,\omega_{\rm r}+\varepsilon(\omega_{\rm q},\omega_{\rm r},|\mathcal A|)\,\tilde{\sigma}_z\,,
\end{equation}
where $\tilde{ \mathbf{n}}=\sum_nn|u_j,n\rangle\rangle\langle\langle n,u_j|$ explicitly enumerates the equivalent replicas and $\tilde{\sigma}_z=\sum_{n,\sigma=\pm}\sigma\ket{\left.u_\sigma,n\right\rangle } \bra{\left\langle u_\sigma,n\right. }$ is the $\sigma_z$ Pauli  matrix acting now diagonally on the Floquet eigenstates. As all replicas are equivalent, we can project onto replica $n=0$, which leads to Eq.~(4) in the main text
\begin{align}\label{eq:diag}
\mathcal H_{\mathcal F}=\varepsilon(\omega_{\rm q}, \omega_{\rm r}, \vert \mathcal A \vert)\,\tilde\sigma_z\,,
\end{align}
where $\tilde{\sigma}_z=\sum_{\sigma=\pm}\sigma\ket{\left.u_\sigma,0\right\rangle } \bra{\left\langle u_\sigma,0\right. }$. 


We can now rely on the expression $\mathcal A=A_{\rm q}+2g_\perp\alpha$, valid when we assume the cavity to be in a coherent state $\ket\alpha$, such that $a\ket\alpha=\alpha\ket\alpha$, to make an expansion of Eq.~\eqref{eq:diag} around $\mathcal A=A_{\rm q}$. Up to second order in $g_\perp \alpha$, the Taylor expansion of the quasi-energy $\varepsilon$ has the general form 
\begin{equation}\label{eq:expansion}
\varepsilon(|A|)=\varepsilon(|A_{\rm q}|)+2g_\perp\alpha\left.\frac{\partial\varepsilon}{\partial\mathcal A}\right|_{A_{\rm q}}+2g_\perp\alpha^*\left.\frac{\partial\varepsilon}{\partial\mathcal A^*}\right|_{A_{\rm q}}+2g_\perp^2\left[\alpha^2\frac{\partial^2\varepsilon}{\partial \mathcal A^2}+2|\alpha|^2\frac{\partial^2\varepsilon}{\partial\mathcal A\partial \mathcal A^*}+\alpha^{*2}\frac{\partial^2\varepsilon}{\partial \mathcal A^{*2}}\right]_{A_{\rm q}}\,.
\end{equation}
A more compact expression is derived by expressing the derivatives along the real and imaginary part of $\mathcal A$, namely
\begin{align}\label{eq:wirtinger_derivatives}
\frac{\partial}{\partial \mathcal{A}}&=\frac{1}{2}\left(\frac{\partial}{\partial \text{Re} \mathcal{A}} - i\frac{\partial}{\partial  \text{Im} \mathcal{A}} \right)\,,&\frac{\partial}{\partial \mathcal{A}^*}&=\frac{1}{2}\left(\frac{\partial}{\partial \text{Re} \mathcal{A}} + i\frac{\partial}{\partial  \text{Im} \mathcal{A}} \right)\,. 
\end{align}
As the derivatives are evaluated on the real axis $\mathcal A=A_{\rm q}$, the first derivative on the imaginary part vanishes 
\begin{equation}\label{eq:Im0}
\left.\frac{\partial \varepsilon}{\partial\mbox{Im}\mathcal A}\right|_{A_{\rm q}}=0\,,
\end{equation}
and we thus find that the linear correction in Eq.~\eqref{eq:expansion} to the Floquet energy $\varepsilon$ close to $\mathcal A=A_{\rm q}$ reads
\begin{align}\label{eq:Sgpar0}
2g_\perp\alpha\left.\frac{\partial\varepsilon}{\partial\mathcal A}\right|_{A_{\rm q}}+2g_\perp\alpha^*\left.\frac{\partial\varepsilon}{\partial\mathcal A^*}\right|_{A_{\rm q}}&=g_\parallel(\alpha+\alpha^*)\,, &  g_\parallel&=g_\perp\left.\frac{\partial \varepsilon}{\partial \mbox{Re}\mathcal A}\right|_{A_{\rm q}}\,,
\end{align}
which coincides with the expression of the longitudinal coupling $g_\parallel$ given in Eq.~(6) in the main text. 

Next to leading order, we neglect the Kerr terms $\propto \alpha^2,\alpha^{*2}$ in Eq.~\eqref{eq:expansion}. One can further apply  the property~\eqref{eq:Im0}  to show 
\begin{align}\label{eq:Schi}
4g_\perp^2|\alpha|^2\left.\frac{\partial^2\varepsilon}{\partial \mathcal A\partial\mathcal A^*}\right|_{A_{\rm q}} &=\chi|\alpha|^2\,, & \chi=g_\perp^2\left[\frac{\partial ^2\varepsilon}{\partial|\mathcal A|^2}+\frac1{|\mathcal A|}\frac{\partial \varepsilon}{\partial |\mathcal A|}\right]_{A_{\rm q}}\,,
\end{align}
which completes our derivation of the dispersive coupling $\chi$ given in Eq.~(6) in the main text.

\subsection{Derivation of Eq.~(7) in the main text and connection with the curvature of the static spectrum}

In the Floquet Hamiltonian~\eqref{eq:Floquet_H}, the terms proportional to $\mathcal A^{(*)}$ are proportional to $\sigma_x$ and $f_\pm$. They thus flip the spin and change the replica index. As a consequence,  the usual perturbative expansion of the system eigen-energies features only even powers of the renormalized drive amplitude $|\mathcal A|$, namely 
\begin{equation}
\varepsilon=\sum_{n=0}^\infty q_n|\mathcal A|^{2n}\,,
\end{equation}
where $q_n$ are coefficients which we do not need to specify at this stage. We are interested then in evaluating the derivatives in Eqs.~\eqref{eq:Sgpar0} and~\eqref{eq:Schi} to leading order in $A_{\rm q}$. One thus finds that 
\begin{align}
\lim_{|\mathcal A|\rightarrow0}\frac{\partial ^2\varepsilon}{\partial|\mathcal A|^2}&=\lim_{|\mathcal A|\rightarrow0}\frac1{|\mathcal A|}\frac{\partial \varepsilon}{\partial |\mathcal A|}=2q_1\,, &&\Longrightarrow &\chi^{(0)}&=4
q_1g_\perp^2=2g_\perp^2\left.\frac{\partial^2\varepsilon}{\partial|\mathcal A|^2}\right|_{A_{\rm q }=0}\,,
\end{align}
which is the expression given for $\chi^{(0)}$ in Eq.~(7) in the main text. Similarly, for the longitudinal coupling we find to linear order in $A_q$
\begin{align}\label{eq:gpchi_small_Aq_SM}
g_{\parallel}^{(0)}&=g_\perp q_1\left.\frac{\partial (\mbox{Re}\mathcal A^2+\mbox{Im}\mathcal A^2)}{\partial \mbox{Re}\mathcal A}\right|_{A_{\rm q}}=2g_\perp q_1 A_{\rm q}=\frac{\chi^{(0)}}{2g_\perp}A_{\rm q}\,,
\end{align}
which completes our derivation of Eq.~(7) in the main text.

We are equipped now to connect  with the notion of curvature and quantum capacitance for both the dispersive~\cite{park_from_adiabatic_2020} and longitudinal coupling~\cite{didier_fast_2015,ruskov2019quantum,ruskov2021modulated,bosco2022fully,michal2023tunable,ruskov2023longitudinal,kohler2017dispersive,kohler2018dispersive}.  The standard 
expression for the dispersive coupling $\chi^{(0)}$ is readily found by relying on standard $2^{\rm nd}$ order perturbation theory. Assuming that the resonator and the qubit are not resonant ($\omega_{\rm r}\neq\omega_{\rm q}$), $g_\parallel$ always vanishes in the $A_{\rm q}\rightarrow0$ limit. Thus, the expansion in $g_\perp \alpha$ of the Floquet quasi-energy~\eqref{eq:diag1} reads $\varepsilon=\omega_{\rm q}/2+\varepsilon^{(2)}$, where
\begin{align}\label{eq:chi0}
\varepsilon^{(2)}&=g_\perp^2\,\langle\langle 0,+|\big(\alpha f_-+\alpha^*f_+\big)\sigma_x\,\frac1{\mathcal H_{\mathcal F,\mathcal A=0}-\frac{\omega_{\rm q}}2}\big(\alpha f_-+\alpha^*f_+\big)\sigma_x\,|+,0\rangle\rangle=\chi^{(0)}|\alpha|^2\,, & \chi^{(0)}&=2g_\perp^2\frac{\omega_{\rm q}}{\omega_{\rm q}^2-\omega_{\rm r}^2}\,,
\end{align}
where we have introduced the eigenstates $|\pm,0\rangle\rangle$ of the undriven problem $\mathcal H_{\mathcal F,\mathcal A=0}$ in the replica $n=0$. The expression of $\chi^{(0)}$ in Eq.~\eqref{eq:chi0} is reproduced in the main text. In the adiabatic limit ($\omega_{\rm{r}}\rightarrow 0$), one thus finds
\begin{align}
\lim_{\omega_{\rm r}\rightarrow0} \chi^{(0)}&=\frac{2g_\perp^2}{\omega_{\rm q}}\,.
\end{align}  
One can readily verify that $1/\omega_{\rm q}$ corresponds to the curvature of the charge qubit at its sweet spot. If we introduce a static detuning $\delta\varepsilon$, the charge qubit Hamiltonian reads $H=\omega_{\rm  q}\,\sigma_z/2 +\delta\varepsilon\,\sigma_x$. Its spectrum is thus given by $\pm\sqrt{\omega_{\rm q}^2+4\delta\varepsilon^2}/2$. Close to the sweet spot  at zero detuning ($\delta\varepsilon=0$), one finds a spectrum $\pm(\omega_{\rm q}/2+\delta\varepsilon^2/\omega_{\rm q})$, where $1/\omega_{\rm q}$ controls the curvature, as expected.

\section{Derivation of the effective Hamiltonian~(8) in the main text}
\label{sec:SW}
In this section, we provide the details about the  derivation of the effective Hamiltonian Eq.~(8), which extends the above results beyond the assumption of a cavity in a coherent state $\ket\alpha$, eigenstate of the photon annihilation operator $a\ket\alpha=\alpha\ket\alpha$. We  start by making the substitutions $\mathcal A\rightarrow A_{\rm q}+2g_\perp\,a$ and $\mathcal A^*\rightarrow \mathcal A^\dagger$ in the time-periodic and Floquet Hamiltonians~\eqref{eq:ham_sup} and~\eqref{eq:Floquet_H}. In particular $\mathcal H_{\mathcal F}$ becomes 
\begin{equation}\label{eq:ham_sup_op}
\bar{\mathcal{H}}_{\mathcal F} = \mathbf{p}\,\omega_{\rm r}+\frac{\omega_{\rm q}}{2}\sigma_z+\frac{1}{2}\left( \mathcal A\,f_{-}+\mathcal A^\dagger\,f_{+} \right)\sigma_x\,.
\end{equation}
To keep notations as light as possible, we have introduced the bar to distinguish $\bar{\mathcal H}_{\mathcal F}$ from $\mathcal H_{\mathcal F}$, where $\mathcal A$ will be considered as an operator or as a complex number respectively and without ambiguity. To make appear the relations~(\ref{eq:Sgpar0}-\ref{eq:Schi}) (Eq.~(6) in the main text), which connect the derivatives of the Floquet spectrum to $g_\parallel$ and $\chi$, it is useful to make explicit the presence of derivatives in $\bar{\mathcal H}_F$, namely
\begin{equation}
\bar{\mathcal{H}}_{\mathcal F} = \mathcal{H}_{\mathcal F}\vert_{\mathcal A=A_{\rm q}}+  2g_\perp a\,\left. \frac{\partial\mathcal{H}_{\mathcal F}}{\partial\mathcal{A}} \right\vert_{\mathcal A=A_{\rm q}} +2g_\perp a^\dagger\,\left. \frac{\partial\mathcal{H}_{\mathcal F}}{\partial\mathcal{A}^*} \right\vert_{\mathcal A=A_{\rm q}} \,.
\end{equation}
To make appear the derivatives of the spectrum, $\mathcal H_{\mathcal F}$ has to be cast in diagonal form, according to the procedure discussed in Section~\ref{sec:numerical_spectrum}. We thus apply to $\bar{\mathcal{H}}_{\mathcal F}$ the unitary transformation $\mathcal{U}$ that diagonalizes $\mathcal H_{\mathcal F}$. The rotated Hamiltonian $\bar{\mathcal H}'_{\mathcal F}=\mathcal{U}\bar{\mathcal{H}}_{\mathcal F}\mathcal{U^\dagger}$ splits in three terms $\bar{\mathcal H}'_{\mathcal F}=\mathcal{D}+\mathcal{H}_1+\mathcal{H}_2$, namely (we drop the notation $|_{\mathcal A=A_{\rm q}}$ to compactify the expressions) 
\begin{align}\label{eq:H_F_decomposition}
\mathcal{D}&=\mathbf{n}\,\omega_{\rm r}+\varepsilon(\omega_{\rm q}, \omega_{\rm r}, \vert \mathcal A \vert)\sigma_z\,, & \mathcal{H}_1&=2g_\perp\left(a\frac{\partial\mathcal D}{\partial\mathcal{A}} + a^\dagger\frac{\partial\mathcal D}{\partial\mathcal{A}^*} \right)\,,& \mathcal{H}_2 &=2g_\perp a\left(\mathcal U \frac{\partial\mathcal U^\dagger}{\partial\mathcal A}\mathcal D \mathcal  + \mathcal D \frac{\partial\mathcal U}{\partial\mathcal A}\mathcal U^\dagger\right)+\mbox{h.c.}\,
\end{align}
where h.c. stands for Hermitian conjugate. The contribution $\mathcal D$ is diagonal in the replica, qubit and Fock basis of the cavity, $\mathcal H_1$ is off-diagonal only in the Fock basis and $\mathcal H_2$ is off-diagonal in all the bases. 

At this stage,  one can immediately show that $\mathcal H_1$  extends the result of Eq.~\eqref{eq:Sgpar0}, without assuming the cavity to be in a coherent state
\begin{align}\label{eq:Sgpar}
\mathcal H_1&=g_\parallel\big( a+a^\dagger)\,,   &  g_\parallel&=g_\perp\left.\frac{\partial \varepsilon}{\partial \mbox{Re}\mathcal A}\right|_{A_{\rm q}}\,.
\end{align}

The off-diagonal term $\mathcal H_2$ in Eq.~\eqref{eq:H_F_decomposition} is the one leading to the dispersive contribution corresponding to Eq.~\eqref{eq:Schi} derived for coherent cavity states in Section~\ref{sec:numerical_spectrum}. To perform the equivalent of the Taylor expansion in $g_\perp \alpha$ of the Floquet spectrum (see Section~\ref{sec:numerical_spectrum}), we perform a Schrieffer-Wolff~\cite{schrieffer1966relation} transformation to $\bar{\mathcal H}'_{\mathcal F }$ up to second order in $g_\perp$. To do so, we introduce the rotated Hamiltonian
\begin{equation}
\bar{\mathcal H}''_{\mathcal F}=e^{i\mathcal S}\bar{\mathcal H}'_{\mathcal F} e^{-i\mathcal S}\simeq \mathcal D+\mathcal H_{1}+\mathcal H_2+i[\mathcal S,\mathcal D]+i[\mathcal S,\mathcal H_{1}+\mathcal{H}_2]-\frac12[\mathcal S,[\mathcal S,\mathcal D]]+\ldots
\label{eq:SW_long}
\end{equation}
The operator $\mathcal{S}$ is derived by imposing the elimination of $\mathcal H_2$
\begin{equation}
i\left[ \mathcal{S},\mathcal D \right]=-\mathcal{H}_2\,,
\label{eq:H_2}
\end{equation}
and is given by
\begin{align}\label{eq:Sgamma}
i\mathcal S&=2g_\perp(\gamma_1 a+\gamma_2 a^\dagger)\,, &\mbox{with}&&\gamma_1 &= -\mathcal{U}\left.\frac{\partial \mathcal U^\dagger}{\partial \mathcal{A}}\right.\,, & \gamma_2 &= -\mathcal{U}\left.\frac{\partial\mathcal{U}^\dagger}{\partial \mathcal{A}^*}\right.=-\gamma_1^\dagger\,.
\end{align}
In these notations, we can write also $\mathcal H_2=-2g_\perp [\gamma_1 ,\mathcal D]a+\mbox{h.c.}$.
Plugging Eq.~\eqref{eq:H_2} into Eq.~\eqref{eq:SW_long} one sees that the second-order corrections generated by $\mathcal{S}$ read 
\begin{equation}
i\left[\mathcal S,\mathcal H_1\right]+\frac i2[\mathcal S,\mathcal H_2]=4g_\perp^2 \left(\left[\gamma_{2}a^\dagger,\frac{\partial\mathcal{D}}{\partial \mathcal{A}}a\right]-\frac{1}{2}\left[\gamma_{2}a^\dagger,\left[\gamma_{1},\mathcal{D}\right]a\right]+\mbox{h.c.} \right)\,,
\label{eq:S2prime}
\end{equation}
where we have neglected all the Kerr terms proportional to $a^2,\,a^{\dagger2}$, which contribute at higher orders. According to Eq.~\eqref{eq:Schi}, our aim is to make apparent the presence of the second derivatives of the spectrum in the dispersive term. Thus, we first establish the connection between such derivatives and the $\gamma_{1,2}$ operator factors introduced in Eq.~\eqref{eq:Sgamma}. From the first derivative
\begin{equation}
\frac{\partial \mathcal D}{\partial \mathcal A}=[\gamma_1,\mathcal D]+\mathcal U\frac{\partial \mathcal H_{\mathcal F}}{\partial \mathcal A}\mathcal U^\dagger\,,
\end{equation}
we construct the second derivative as
\begin{align}
\frac{1}{2}\frac{\partial^2 \mathcal D}{\partial \mathcal{A}\partial\mathcal{A}^*}&=\frac12 \frac{\partial}{\partial \mathcal A^*}\left[\gamma_1,\mathcal{D}\right]+\frac{1}{2}\left[\gamma_2, \mathcal U\frac{\partial \mathcal H_{\mathcal F}}{\partial \mathcal A}\mathcal U^\dagger \right] \nonumber\\
&=\frac{1}{2}\frac{\partial}{\partial \mathcal{A}^*}\left[\gamma_1,\mathcal{D}\right]+\frac12 \left[\gamma_{2},\frac{\partial\mathcal{D}}{\partial \mathcal{A}}\right]-\frac{1}{2}\left[\gamma_{2},\left[\gamma_{1},\mathcal{D}\right]\right]\,.
\end{align}
Then, making use of the identities $[AB,C]=A[B,C]+[B,C]A$ and $AB=\{ B,A\}/2-[B,A]/2$, we rearrange the terms in Eq.~\eqref{eq:S2prime} as
\begin{align}\label{eq:anti-comm}
i\left[\mathcal S,\mathcal H_1\right]+\frac i2[\mathcal S,\mathcal H_2]&=\chi \left(a^\dagger a+\frac12\right)\sigma_z+\mathcal V\,, & \chi=4g_\perp^2\left.\frac{\partial^2 \varepsilon}{\partial \mathcal A\partial \mathcal A^*}\right|_{A_{\rm q}}=g^2_\perp\left[ \frac{\partial^2 \varepsilon}{\partial |\mathcal A |^2} + \frac{1}{|\mathcal A|} \frac{\partial \varepsilon}{\partial | \mathcal A |} \right]_{A_{\rm q}}\,,
\end{align}
where we have introduced the term leading to Eq.~(8) in the main text plus an additional one
\begin{align}
\mathcal V&=2g_\perp^2\left(\left[\frac{\partial\gamma_2}{\partial \mathcal{A}},\mathcal{D}\right]aa^\dagger-\frac{1}{2}\left[ \gamma_2,\frac{\partial \mathcal D}{\partial \mathcal A}\right]-\left\{\gamma_2,\frac{\partial\mathcal{D}}{\partial\mathcal{A}}\right\}+\frac12 \left\{\gamma_2,\left[ \gamma_1, \mathcal D\right] \right\}+\text{h.c.} \right)\,.
\label{eq:V}
\end{align}
We show now that $\mathcal V$ is either off-diagonal in the qubit and replica space, either is proportional to the identity operator in both the qubit and the replica space. Since all the off-diagonal terms can be eliminated at higher orders, they can be here neglected when projecting into the diagonal subspace. 

The first two terms in $\mathcal V$ are off-diagonal, since they are commutators with a diagonal operator. To discuss the other terms, we inspect the general structure of the operators $\gamma_{1,2}$. Using Eq.~\eqref{eq:floquet_notation}, we write explicitly the operator $\mathcal U^\dagger$, whose columns are the eigenvectors of $\mathcal H_{\mathcal F}$  
\begin{align}
\mathcal U^\dagger=\sum_{jn} \ket{\left.u_{j},n\right\rangle } \bra{j}\left( n \right\vert=\sum_{jn}f^{-n} \ket{u_{j,n}}\bra{j}\,,
\label{eq:matrix_U}
\end{align}
such that $\gamma_1$ reads
\begin{align}
\gamma_1 &= -\mathcal U \frac{\partial \mathcal U^\dagger}{\partial \mathcal A}=-\sum_{jj'\alpha} \Gamma_{jj'}^\alpha f^\alpha \ket{j}\bra{j'}\,, & \Gamma_{jj'}^\alpha = \sum_n \bra{u_{j,n}}\left(\frac{\partial}{\partial \mathcal A} \ket{u_{j',n-\alpha}} \right)\,,
\end{align}
where we have introduced the notation $f^\alpha \equiv f^{\vert \alpha \vert} _{\text{sgn}\alpha}$. The coefficients of $\gamma_2=-\gamma_1^\dagger$ are related to the ones of $\gamma_1$ through the important property $\Gamma_{jj'}^{\alpha*}=-\Gamma_{1-j,1-j'}^\alpha$, that we will prove below.
Considering the symmetry transformation 
\begin{align}
\mathcal{W}&=\sum_n (-1)^n \vert -n ) ( n \vert \sigma_x\,, &
\mathcal{W}\mathcal{H}_{\mathcal F}\mathcal{W}^\dagger &= -\mathcal{H}_{\mathcal F}^{*}\,,
\end{align}
where $*$ takes the complex conjugation of the coefficients buried in the operator ($\mathcal A \longleftrightarrow \mathcal A^*$), given one of the two eigenvectors in the zero-th replica (or in any other replica subspace) we can construct the other one as $\vert u_{1-j}, 0 \rangle \rangle = \mathcal W \vert u_j, 0 \rangle \rangle^*$. Indeed
\begin{equation}
 \mathcal{H}_{\mathcal F} \mathcal{W} \vert u_j,0 \rangle \rangle^*=(\mathcal{H}_{\mathcal F}^* \mathcal{W} \vert u_j,0 \rangle \rangle)^*=-\varepsilon_j \mathcal{W}\vert u_j,0 \rangle \rangle^*\,,
\end{equation}
which, since $\varepsilon_{1-j}=-\varepsilon_j$, proofs that $\mathcal W \vert u_j, 0 \rangle \rangle^*=\vert u_{1-j}, 0 \rangle \rangle$. We remark that $*$ only takes the complex conjugate of the coefficients, and thus the meaning of $\mathcal W \vert u_j, 0 \rangle \rangle^*=\vert u_{1-j}, 0 \rangle \rangle$ is 
\begin{equation}
u_{j,n,\sigma}=(-1)^n u_{1-j,-n,1-\sigma}^*\,.
\end{equation}
Finally, we have $(\partial_{\mathcal A}\ket{u_{j,n}})^*=\mathcal{W}\partial_{\mathcal A^*}\ket{u_{1-j,n}}$, hence $\Gamma_{jj'}^{\alpha*}=-\Gamma_{1-j,1-j'}^\alpha$.
We consider now the first anti-commutator in Eq.~\eqref{eq:V}
\begin{equation}
-\left\{ \gamma_2,\frac{\partial \mathcal D}{\partial \mathcal A} \right\} + \text{h.c.} = -\frac{\partial \varepsilon}{\partial \mathcal A}\sum_{\alpha} f^\alpha (\Gamma_{11}^{\alpha*}\ket{1}\bra{1}-\Gamma_{00}^{\alpha*}\ket{0}\bra{0}) +\text{h.c.}\,.
\end{equation}
Since the derivatives are taken on the real axis, we have $\partial_{\mathcal A}\varepsilon=\partial_{\mathcal A^*}\varepsilon=\partial_{\text{Re}\mathcal A}\varepsilon/2$ (see Eqs.~\eqref{eq:wirtinger_derivatives}, \eqref{eq:Im0}), and the only non off-diagonal term is $(\partial_{\text{Re}\mathcal A}\varepsilon/2)(\Gamma_{00}^0-\Gamma_{11}^0)\mathds{1}$, that is proportional to the identity operator. For the second anti-commutators, we have
\begin{align}
\left\{ \gamma_2 \left[ \gamma_1, \mathcal D \right] \right\} + \text{h.c.} &= -\sum_{jj'k\alpha\beta} f^{\alpha-\beta} \Gamma_{jj'}^{\alpha} \left( \Gamma_{jk}^{\beta*}\ket{k}\bra{j'}+\Gamma_{kj'} \ket{j}\bra{k} \right)(\varepsilon_{j'}-\varepsilon_{j}-\alpha\omega_{\rm r}) + \text{h.c.}\,,
\end{align}
whose only non off-diagonal terms are
\begin{align}
-\sum_{jj'\alpha} \Gamma_{jj'}^{\alpha} \Gamma_{jj'}^{\alpha*}\left(\ket{j'}\bra{j'}+ \ket{j}\bra{j}\right)(\varepsilon_{j'}-\varepsilon_{j}-\alpha\omega_{\rm r}) + \text{h.c.}=-4\sum_\alpha\alpha\omega_{\rm r}(\Gamma_{01}^\alpha\Gamma_{10}^\alpha +\Gamma_{00}^\alpha\Gamma_{11}^\alpha)\mathds{1} + \text{h.c.}\,,
\end{align}
that are again only proportional to the identity operator.
We have shown that all the terms in Eq.~\eqref{eq:V} are either off-diagonal in the replica or qubit space, or proportional to the identity operator, and thus they can be neglected in the Hamiltonian after projection on the diagonal subspace.

So, projecting to the diagonal subspace up to order $g_\perp^2$, we find  the effective  Hamiltonian 
\begin{equation}
\mathcal H_{\mathcal F}''=e^{i\mathcal S}\mathcal{U}\bar{\mathcal H}_{\mathcal F}\mathcal{U^\dagger} e^{-i\mathcal S} = \mathcal{D}+g_\parallel(a+a^\dagger)\sigma_z+\chi \left(a^\dagger a+\frac{1}{2}\right)\sigma_z\,.
\end{equation}
Inverting the rotation, we can write $\bar{\mathcal{H}}_{\mathcal F}$ as
\begin{equation}\label{eq:hbarfinal}
\bar{\mathcal{H}}_{\mathcal F} =\tilde{\mathbf{n}} \omega_{r} + \varepsilon(\omega_{\rm q}, \omega_{\rm r}, \vert \mathcal A \vert) \tilde{\sigma}_z+g_\parallel(\tilde a+\tilde a^\dagger)\tilde\sigma_z+\chi\left(\tilde a^\dagger \tilde a+\frac{1}{2}\right)\tilde \sigma_z\,,
\end{equation}
where we switched to the operators $\tilde {\mathcal O}=\mathcal U^\dagger e^{-i\mathcal S}\mathcal Oe^{i\mathcal S}\mathcal U$, which are diagonal in the basis of the eigenvectors of the system. As shown in Eqs.~\eqref{eq:Sgpar}~and~\eqref{eq:anti-comm}, the longitudinal and dispersive couplings,  $g_\parallel$ and $\chi$, coincide with those given by Eq.~(6) in the main text 
\begin{align}\label{eq:gperpchi_sup}
g_\parallel&= g_\perp \left. \frac{\partial \varepsilon}{\partial  \mbox{Re} \mathcal A } \right\vert_{A_{\rm q}},&
\chi &= g^2_\perp\left[ \frac{\partial^2 \varepsilon}{\partial |\mathcal A |^2} + \frac{1}{|\mathcal A|} \frac{\partial \varepsilon}{\partial | \mathcal A |} \right]_{A_{\rm q}}\,.
\end{align}
The projection of the Floquet Hamiltonian into the zero-th replica space gives
 Eq.~(8) in the main text, where we omitted the $\sim$ symbol on the $a,\,a^\dagger$ operators for simplicity. 

\section{Extensions: additional drive to the cavity, dispersive readout and Purcell effect}
\label{sec:additional_drive}
The study of realistic multi-level systems and the comparison to dispersive readout requires the extension of the previous results to the case in which both the qubit and the resonator are driven at a generic frequency $\omega_{\rm d}$ and the cavity drive is phase-shifted by a phase $\varphi$. We thus consider an  Hamiltonian of the form 
\begin{equation}\label{eq:H_additional_drive}
\mathcal{H}=\frac{\omega_{\rm q}}{2}\sigma_z +  g_\perp (a+a^\dagger)\sigma_x + \omega_{\rm r}a^\dagger a+A_{\rm q}\cos(\omega_{\rm d}t)\sigma_x+A_{\rm r}\cos(\omega_{\rm d}t+\varphi)(a+a^\dagger)\,.
\end{equation} 

In the rotating frame defined by $\mathcal W(t)=e^{i\omega_{\rm d}t a^\dagger a}$, the rotated Hamiltonian $\mathcal H'=\mathcal W\mathcal H\mathcal W^\dagger+i(\partial_t \mathcal W)\mathcal W^\dagger$ reads
\begin{equation}
\begin{aligned}
\mathcal{H}^\prime =\frac{\omega_{\rm q}}{2}\sigma_z + \frac12\left(\mathcal Ae^{-i\omega_{\rm d} t}+\mathcal A^\dagger e^{i\omega_{\rm d} t}\right)\sigma_x+\frac{A_{\rm r}}{2}\left( a e^{\varphi}+a^\dagger e^{-i\varphi}\right)+(\omega_{\rm r}-\omega_{\rm d})a^\dagger a\,,
\end{aligned}
\end{equation}
where we neglect the fast terms rotating at $\pm 2\omega_{\rm d}t$. 
Following the same procedure used in Section~\ref{sec:SW}, one finds the extended version of Eq.~\eqref{eq:hbarfinal}, namely
\begin{equation}\label{eq:H_floquet_additional_drive}
\begin{aligned}
\bar{ \mathcal{H}}_{\mathcal F}&=\tilde{\mathbf{n}}\,\omega_{\rm d}+\varepsilon(\omega_{\rm q}, \omega_{\rm d}, \vert \mathcal A \vert)\tilde\sigma_z+g_\parallel(\tilde a+\tilde a^\dagger)+\chi\left(\tilde a^\dagger \tilde a+\frac12\right)\tilde\sigma_z+\frac{A_{\rm r}}{2}\left( \tilde a e^{i\varphi}+\tilde a^\dagger e^{-i\varphi} \right)+(\omega_{\rm r}-\omega_{\rm d})\tilde a^\dagger \tilde a\,.
\end{aligned}
\end{equation}
Standard dispersive readout is achieved by sending 
$A_{\rm q}=0$ and $\omega_{\rm d}=\omega_{\rm r}$ in the above expression. Choosing $\varphi=3\pi/2$ to ease comparison with the lonigitudinal readout protocol in the main text, the effective Hamiltonian reads~\cite{blais_cQED_2004}
\begin{equation}
\mathcal{H}_{\mathcal F}=\varepsilon(\omega_{\rm q}, \omega_{\rm r}, \vert \mathcal A=0 \vert)\tilde\sigma_z+\left.\chi\right\vert_{\mathcal A=0}\left(\tilde a^\dagger \tilde a+\frac12\right)\tilde\sigma_z+\frac{A_{\rm r}}{2i}\left( \tilde a - \tilde a^\dagger  \right)\,,
\end{equation}
with $\left.\chi\right\vert_{\mathcal A=0}=\chi^{(0)}$ given in Eq.~\eqref{eq:chi0}.

We also give an alternative derivation of the Purcell effect within our formalism, which is equivalent to standard approaches~\cite{blais_cQED_2004}. We rely on a Schrieffer-Wolff transformation~\cite{schrieffer1966relation} to obtain a perturbative expression of the unitary $U$ at small values of $\mathcal{A}$ to diagonalize $\mathcal D$ in Eq.~\eqref{eq:H_F_decomposition}. Similarly to the Schrieffer-Wolff procedure described in Section~\ref{sec:SW} we split the Floquet Hamiltonian in a diagonal and off-diagonal term
\begin{align}
\mathcal H_{\mathcal F} &= \mathcal H_{0}+\mathcal V\,, & \mathcal H_0 &= \mathbf{n}\omega_{\rm r}+\frac{\omega_{\rm q}}{2}\sigma_z\,, & \mathcal V &= \frac{1}{2}\left( \mathcal Af_-+\mathcal{A}^*f_+\right)\sigma_x\,,
\end{align}
and look for an operator $\mathcal T$ such that $\mathcal{U}=e^{i\mathcal{T}}$ and satisfying
\begin{equation}
i[\mathcal T,\mathcal H_0]=-\mathcal V\,.
\end{equation}
This equation is solved by 
\begin{equation}\label{eq:perturbative_sol}\begin{aligned}
i\mathcal T&=\frac12\left(\frac{\mathcal{A} }{\omega_{\rm q}-\omega_{\rm r}}f_-+\frac{\mathcal{A}^* }{\omega_{\rm q}+\omega_{\rm r}} f_+  \right) \sigma_+- \frac12\left(\frac{\mathcal{A} }{\omega_{\rm q}+\omega_{\rm r}}f_-+\frac{\mathcal{A}^* }{\omega_{\rm q}-\omega_{\rm r}}f_+\right)\sigma_-\,.
\end{aligned}
\end{equation}
The Purcell effect is a qubit relaxation mediated by emission of photons in the resonator. To estimate this relaxation rate, we consider first the fact that photons in the cavity decay at rate $\kappa$. This decay is described by a Lindblad equation, which describes the evolution of the density matrix $\rho $ of the system
\begin{align}
\frac{\partial\rho}{\partial t}&=-i[\mathcal H,\rho]+\kappa\mathcal D_a[\rho] \,, &\mathcal{D}_a \left[\rho\right]&=a\rho a^{\dagger}-\frac{1}{2}\left\{ a^{\dagger}a,\rho\right\}\,.
\end{align}
However, in the rotated frame derived in Section~\ref{sec:SW},  where the unitary dynamics is diagonal, the dissipative term becomes  $\kappa\mathcal{D}(a)=\kappa\mathcal{D}( e^{i\mathcal S}\mathcal U\tilde a \mathcal U^\dagger e^{-i\mathcal S})$. To do so, we compute $\gamma_1$ up to zero-th order in $\mathcal A$ through Eq.~\eqref{eq:perturbative_sol}
\begin{equation}
\gamma_1 = -\mathcal{U}\frac{\partial \mathcal U^\dagger}{\partial\mathcal{A}}=\frac{\partial \mathcal U}{\partial\mathcal{A}}\mathcal U^\dagger=i\frac{\partial \mathcal T}{\partial  \mathcal{A}}=\frac12\frac{1}{\omega_{\rm q}-\omega_{\rm r}}f_-\sigma_{+}-\frac12\frac1{\omega_{\rm q}+\omega_{\rm r}}f_-\sigma_{-}\,.
\end{equation}
Up to linear order in $g_\perp$ and zero-th order in $\mathcal A$ we have
\begin{equation}\label{eq:atildea}
\begin{aligned}
a \simeq \tilde a + 2g_\perp\gamma_1^\dagger=\tilde a +\frac{g_\perp}{\omega_{\rm q}-\omega_{\rm r}}f_+ \sigma_--\frac{g_\perp }{\omega_{\rm q}+\omega_{\rm r}}f_+\sigma_+\,.
\end{aligned}
\end{equation}
Considering $\sigma_- \simeq \tilde \sigma_-$ and neglecting sub-dominant terms proportional to $g_\perp^2/(\omega_{\rm q}+\omega_{\rm r})^2$, we obtain $\kappa\mathcal{D}(a)\approx\kappa\mathcal{D}(\tilde a)+\gamma\mathcal{D}(\tilde\sigma_-)$, describing an effective decay for the qubit at rate~\cite{blais_cQED_2004} 
\begin{equation}
\gamma=\frac{\kappa g_\perp^2}{(\omega_{\rm q}-\omega_{\rm r})^2}\,.
\label{eq:purcell}
\end{equation}

\section{Derivation of the effective Hamiltonian~(9) for multi-level systems}
\label{sec:generalization_m_levels}
In this section we give a generalization of the Floquet approach of Section~\ref{sec:numerical_spectrum} to multi-level systems, leading to  Eq. (9) in the main text. We consider systems described by a general Hamiltonian $\mathcal{H}_{\rm sys}[Q]$. The Hamiltonian depends on a generic control parameter $Q$, that corresponds to an excess of charge on a capacitor. Charge fluctuations caused by the classical drive and by the resonator lead to $Q\rightarrow Q_0+\delta Q$, with $\delta Q = A_{\rm q}\cos(\omega_{\rm d}t)+g_\perp(a+a^\dagger)$. The coupling $g_{\perp}$, depends on the coupling scheme whereas $\omega_{\rm d}$ is the strength of the drive. To apply the compensation tones described in the main text, we also consider an additional drive to the cavity as in Eq.~\eqref{eq:H_additional_drive}. Expanding the Hamiltonian till linear order in $\delta Q$ we obtain
\begin{align}
\mathcal{H}&=\mathcal{H}_{\rm sys}[Q_0]+\left. \frac{\partial \mathcal H_{\rm sys}}{\partial Q} \right\vert_{Q_0} \delta Q +\omega_{\rm r}a^\dagger a+A_{\rm r}\cos(\omega_{\rm d}t+\varphi)(a+a^\dagger)\,.
\label{eq:multi_level_H_0}
\end{align}

Following the same procedures detailed in Sections~\ref{sec:floquet_theory} and~\ref{sec:additional_drive}, we build the Floquet Hamiltonian 
\begin{equation}
\mathcal{H}_{\mathcal F}=\mathcal{H}_{\rm sys}[Q_0]\otimes \mathds{1}_{\mathcal F}+\frac{1}{2}\left( \mathcal{A} f_{-}+\mathcal A^\dagger f_{+} \right)\left.\frac{\partial \mathcal H_{\rm sys}}{\partial Q}\right\vert_{Q_0}+\frac{A_{\rm r}}{2}\left( \tilde a e^{i\varphi}+\tilde a^\dagger e^{-i\varphi} \right)+(\omega_{\rm r}-\omega_{\rm d})\tilde a^\dagger \tilde a\,,
\label{eq:floquet_H_multi_level}
\end{equation}
that, projected into the coherent states $\ket{\alpha}$ and in the basis set $\{ \vert p ) \}$ of the extended replica space, reads
\begin{equation}
\mathcal H_{\mathcal F}=
\begin{pmatrix}\ldots\\
 & \omega_{\rm r}\mathds{1}+\mathcal{H}_{\rm sys} & \left.\frac{\mathcal A^*}{2}\frac{\partial \mathcal H_{\rm sys}}{\partial Q}\right\vert_{Q_0} & 0 \\
 & \frac{\mathcal A}{2}\left.\frac{\partial \mathcal H_{\rm sys}}{\partial Q}\right\vert_{Q_0}  & \mathcal{H}_{\rm sys} & \frac{\mathcal A^*}{2}\left.\frac{\partial \mathcal H_{\rm sys}}{\partial Q}\right\vert_{Q_0} \\
 & 0  & \frac{\mathcal A}{2}\left.\frac{\partial \mathcal H_{\rm sys}}{\partial Q}\right\vert_{Q_0} & -\omega_{\rm r}\mathds{1}+\mathcal{H}_{\rm sys} \\
 &  &  &  &    \ldots
\end{pmatrix}\,,
\label{eq:projected_floquet_H_multi_level}
\end{equation}
where $\mathds{1}$ is the identity acting on the Hilbert space spanned by the system Hamiltonian $\mathcal H_{\rm sys}$. 
The problem  can be in principle diagonalized leading to the quasi-energies $\varepsilon_j$. We restrain to the zero replica. Assuming that the Schrieffer-Wolff trasnformation discussed in Section~\ref{sec:SW} can be extended to this case, the Taylor expansion of the quasi-energies around $\mathcal A=A_{\rm q}$ leads to the effective Hamiltonian
\begin{equation}
\mathcal{H}_{\mathcal F}=\sum_j \left[ \varepsilon_j(\omega_{\rm q},\omega_{\rm d}, \vert A_{\rm q} \vert)+\frac{1}{2}\chi_j \right] \tilde{\mathcal{P}}_j + g_{\parallel,j}(\tilde a+\tilde a^\dagger) \tilde{\mathcal{P}}_j + \chi_j \tilde a^\dagger \tilde a \tilde{\mathcal{P}}_j +\frac{A_{\rm r}}{2}\left( \tilde a e^{i\varphi}+\tilde a^\dagger e^{-i\varphi} \right)+(\omega_{\rm r}-\omega_{\rm d})\tilde a^\dagger \tilde a\,,
\label{eq:effective_H_beyond}
\end{equation}
where we have introduced the projectors over the $j$-th Floquet eigenmodes $\tilde{\mathcal{P}}_j=\ket{u_j(0)}\bra{u_j(0)}$. Notice the presence of the term $\chi_j\tilde{\mathcal P}_j/2$, which  accounts for quantum effects in the cavity. Each Floquet mode thus couples longitudinally and dispersively to the cavity independently. The expressions of $g_{\parallel,j}$ and $\chi_j$ are given by
\begin{align}
g_{\parallel,j} &= g_\perp \left. \frac{\partial \varepsilon_j}{\partial \text{Re} \mathcal A } \right\vert_{A_{\rm q}}, &
\chi_j &= g^2_\perp\left[\frac{\partial^2 \varepsilon_j}{\partial |\mathcal A |^2} + \frac{1}{|\mathcal A|} \frac{\partial \varepsilon_j}{\partial | \mathcal A |} \right]_{A_{\rm q}}\,.
\label{eq:g_parallel_chi}
\end{align}
A projection of Eq.~\eqref{eq:effective_H_beyond} onto the manifold spanned by the two lowest levels $\ket{0}$ and $ \ket{1}$ leads to 
\begin{equation}
\begin{aligned}
\mathcal{H}_{\mathcal F} = \left(\varepsilon(A_{\rm q}) + \frac{\chi}{2} \right)\tilde\sigma_z+ g_\parallel(\tilde a+\tilde a^\dagger)\tilde\sigma_z + \chi \tilde{a}^\dagger \tilde a\tilde\sigma_z + \bar{g}_\parallel (\tilde a+\tilde a^\dagger) + \bar{\chi}\tilde a^\dagger \tilde a +\frac{A_{\rm r}}{2}\left( \tilde a e^{i\varphi}+\tilde a^\dagger e^{-i\varphi} \right)+(\omega_{\rm r}-\omega_{\rm d})\tilde a^\dagger \tilde a\,,
\label{eq:not_compensated}
\end{aligned}
\end{equation}
where we have mapped the system onto a qubit Hamiltonian whose energy is given by 
\begin{equation}
\varepsilon(A_{\rm q})=\frac12\Big[\varepsilon_1(A_{\rm q})-\varepsilon_0(A_{\rm q})\Big]\,,
\end{equation}
and with the resulting couplings
\begin{align}
g_\parallel &=\frac12\left(g_{\parallel,1}-g_{\parallel,0}\right)\,,
&\bar{g}_\parallel&=\frac12\left(g_{\parallel,1}+g_{\parallel,0}\right)\,, & \chi&=\frac12(\chi_1-\chi_0)\,, & \bar{\chi}&=\frac12(\chi_1+\chi_0)\,.
\label{eq:gparallel_chi_multilevel}
\end{align}
As we will discuss in detail in Section~\ref{sec:evolution_a}, the terms proportional to $\bar{\chi}$ and $\bar{g}_\parallel$ in Eq.~\eqref{eq:not_compensated} give rise to a qubit-independent dynamics. In order to compensate these terms we choose $A_{\rm r}=-2\bar{g}_\parallel$, $\varphi=0$ and $\omega_{\rm d}-\omega_{\rm r}=\bar{\chi}$, such that Eq.~\eqref{eq:not_compensated} reproduces the simple form given by Eq. (9) in the main text, the closest leading to a pure longitudinal splitting of the cavity dynamics.

\section{Cavity dynamics calculations. Analytics and numerics. }
In this Section we give details about the analytical and numerical calculations to derive the cavity dynamics and their comparison. Analytics are possible as they are performed starting from the effective model~\eqref{eq:not_compensated}, while numerics simulate exactly the dynamics of the parent model~\eqref{eq:multi_level_H_0}.

\subsection{Analytical integration of the equations of motion of the cavity field expectation value}\label{sec:evolution_a}
Considering the Hamiltonian~\eqref{eq:not_compensated} and the dual expression of the Lindblad operator, the equations of motion for the expectation value of an operator $\mathcal O$ read
\begin{align}\label{eq:heisenberg}
\partial_t \langle\mathcal O\rangle&=i\langle[\mathcal H_{\mathcal F},\mathcal O]\rangle+\kappa\mathcal D^\dagger_{\tilde a}[\mathcal O]+\gamma\mathcal D^\dagger_{\tilde \sigma_-}[\mathcal O] & \mathcal{D}^\dagger_X\left[\langle\mathcal O \rangle\right]&=\langle X^{\dagger}\mathcal O X\rangle -\frac{1}{2}\langle\left\{ X^{\dagger}X,\mathcal O\right\}\rangle\,,
\end{align}
where $\kappa$ is the cavity photon decay rate and $\gamma$ is the Purcell decay rate of the spin given by Eq.~\eqref{eq:purcell}. 
Applying equation~\eqref{eq:heisenberg} to the evolution of $\langle \tilde a\rangle$, one finds that the following set of coupled equations needs to be solved
\begin{equation}\begin{aligned}
\partial_{t}\tilde a&=-i\chi \tilde a\tilde\sigma_{z}-a\left[ i\bar{\chi}+\frac{\kappa}{2}+i(\omega_{\rm r}-\omega_{\rm d}) \right]-ig_\parallel \tilde\sigma_z-i\left(\bar{g}_\parallel+\frac{A_{\rm r}}{2}e^{-i\varphi}\right)\,,\\
\partial_{t}\left[\tilde  a\tilde\sigma_z\right]&=-\left[i\bar{\chi}+\frac{\kappa}{2}+\gamma+i(\omega_{\rm r}-\omega_{\rm d}) \right]\tilde a\tilde\sigma_z-\tilde a(i\chi+\gamma)-ig_\parallel-i\left(\bar{g}_\parallel+\frac{A_{\rm r}}{2}e^{-i\varphi}\right)\tilde\sigma_z\,,\\
\partial_{t}\tilde\sigma_{z}&=-\gamma\left(1+\tilde\sigma_{z}\right)\,.
 \label{eq:system}
\end{aligned}
\end{equation}
The dynamics of $\tilde\sigma_z$ is immediately obtained as
\begin{equation}
\tilde\sigma_{z}\left(t\right)=-1+\left(\tilde\sigma_{z}\left(0\right)+1\right)e^{-\gamma t}\,.
\end{equation}
To solve the system of differential equations we introduce the normal modes 
\begin{align}
\alpha&=\tilde a-\frac{i\chi}{\gamma+i\chi}\tilde a\tilde\sigma_{z}\,,&\beta&=a+a\tilde\sigma_z\,.
\label{eq:new_variables}
\end{align}
The system in Eq.~\eqref{eq:system} can now be decoupled and reads
\begin{equation}
\begin{aligned}
\partial_{t}\alpha & = -\left[i(\bar{\chi}-\chi)+i(\omega_{\rm r}-\omega_{\rm d})+\frac{\kappa}{2}\right]\alpha-i\left[ g_\parallel-\frac{i\chi}{\gamma+i\chi}\left(\bar{g}_\parallel+\frac{A_{\rm r}}{2}e^{-i\varphi}\right) \right]\left(\tilde\sigma_z+1\right)\\&\quad\:+i\frac{\gamma+2i\chi}{\gamma+i\chi}\left(g_\parallel-\bar{g}_\parallel-\frac{A_{\rm r}}{2}e^{-i\varphi}\right),\\
\partial_{t}\beta & =-\left[i(\chi+\bar\chi)+i(\omega_{\rm r}-\omega_{\rm d})+\frac{\kappa}{2}+\gamma \right]\beta-i\left[ g_\parallel+\bar{g}_\parallel+\frac{A_{\rm r}}{2}e^{-i\varphi} \right](\tilde\sigma_z+1).
\label{eq:system_2}
\end{aligned}
\end{equation}
Considering an empty cavity at $t=0$, the solutions of Eq.~\eqref{eq:system_2} are
\begin{align}
\alpha\left(t\right) & =-i\left[\left( \bar{g}_\parallel-g_\parallel + \frac{A_{\rm r}}{2}e^{-i\varphi} \right)\left(\frac{\gamma+2i\chi}{\gamma+i\chi}\right)\frac{1-e^{-\frac{\kappa}{2}t-i(\bar{\chi}-\chi )t-i(\omega_{\rm r}-\omega_{\rm d})t}}{i(\bar{\chi}-\chi)+i(\omega_{\rm r}-\omega_{\rm d})+\kappa/2}\right.\label{eq:alpha}\\ \nonumber
&\quad \left.+\left(\tilde\sigma_z\left(0\right)+1\right)\left( g_\parallel-\frac{i\chi}{\gamma+i\chi} \left( \bar{g}_\parallel+\frac{A_{\rm r}}{2}e^{-i\varphi} \right)\right)\frac{e^{-\gamma t}-e^{-\frac{\kappa}{2}t-i(\bar{\chi}-\chi )t-i(\omega_{\rm r}-\omega_{\rm d})t}}{i(\bar{\chi}-\chi)+i(\omega_{\rm r}-\omega_{\rm d})+\kappa/2-\gamma}\right],\\
\beta\left(t\right) & =-i\left(\tilde\sigma_{z}\left(0\right)+1\right)\left( \bar{g}_\parallel+g_\parallel+\frac{A_{\rm r}}{2}e^{-i\varphi} \right)\frac{1-e^{-\frac{\kappa}{2}t-i(\bar{\chi}+\chi )t-i(\omega_{\rm r}-\omega_{\rm d})t}}{i(\bar{\chi}+\chi)+i(\omega_{\rm r}-\omega_{\rm d})+\kappa/2}e^{-\gamma t}\,,\label{eq:beta}
\end{align}
and the evolution of $\langle a\rangle(t)$ is obtained through
\begin{equation}
\langle a (t)\rangle=\frac{1}{\gamma+2i\chi}\big[\left(\gamma+i\chi\right)\alpha\left(t\right)+i\chi\beta\left(t\right)\big]\,.
\label{eq:cavity_field}
\end{equation}
Based on this analytical formula, we compute the trajectories in Figs. 2, 3, 4 in the main text. In particular, the coupling constants $g_\parallel$, $\bar{g}_\parallel$, $\chi$, $\bar{\chi}$ are computed by taking numerical derivatives of the Floquet spectrum, see Section~\ref{sec:floquet_theory}. To account for the effect of the classical drive on the qubit, we compute $\langle \tilde \sigma_z(0) \rangle$ through $\tilde\sigma_z=\mathcal{U}^\dagger \sigma_z\mathcal{U}$, neglecting the effect of the resonator. We obtain $\mathcal{U}^\dagger$ through Eq.~\eqref{eq:matrix_U}, where the eigenvectors are computed numerically solving the spectrum of Eq.~\eqref{eq:floquet_matrix}. We give below the simpler expressions which describe dispersive and longitudinal readout for an ideal charge qubit. We then discuss the effects of a finite $\bar\chi$ and $\bar g_\parallel$ on longitudinal readout and show how they can be compensated by applying a fine-tuned tone to the cavity.

\subsubsection{Dispersive and longitudinal readout for ideal charge qubit and compensation tones}\label{sec:compensation}
{\it Dispersive readout -- }We consider here the simplest case of an ideal charge qubit where $\bar \chi=\bar g_\parallel=0$ and neglect Purcell effect ($\gamma=0$).  In this case, the standard analytic formula for the evolution of the cavity $\langle a\rangle$ with dispersive readout is obtained by driving only the resonator at the frequency $\omega_{\rm d}=\omega_{\rm r}$. We thus set $A_{\rm q}=0$ and $\varphi=3\pi/2$ to ease comparison with longitudinal readout in Fig. 2 in the main text. Equation~\eqref{eq:cavity_field} simplifies to 
\begin{equation}
\langle a(t)\rangle_{\rm disp.} = \frac{A_{\rm r}}{4}\sum_{\sigma=\pm 1} \sigma \left[ \big( \langle\tilde\sigma_z(0)\rangle + \sigma\big)\frac{1-e^{-\frac{\kappa}{2}t-i\sigma \chi t}}{i\sigma\chi+\frac{\kappa}{2}} \right]\,.
\end{equation}
In the stationary $t\rightarrow\infty$ limit, assuming $\langle\tilde \sigma_z(0)\rangle\simeq\sigma$, one finds that  $\langle a(t=\infty)\rangle_{\rm disp.}=A_{\rm r}/2(i\sigma\chi+\kappa/2)$, while, at short times, the expansion of the exponential gives $\langle a(t\rightarrow 0)\rangle_{\rm disp.}\approx A_{\rm r}t/2$, that is qubit state independent. Both these asymptotic expressions are given in the main text.

{\it Longitudinal readout -- }An analog expression is derived for the longitudinal driving protocol  
\begin{equation}\label{eq:alongchi}
\langle a(t)\rangle_{\rm long. } = -\frac{ig_\parallel}{2}\sum_{\sigma=\pm 1} \left[ ( \langle\tilde\sigma_z(0)\rangle + \sigma)\frac{1-e^{-\frac{\kappa}{2}t-i\sigma \chi t}}{i\sigma\chi+\frac{\kappa}{2}} \right]\,,
\end{equation}
which, if $\chi \approx 0$, can be further simplified to
\begin{equation}
\langle a(t)\rangle_{\rm long. }=-i g_\parallel \langle \tilde\sigma_z(0)\rangle\left( \frac{1-e^{-\frac{\kappa}{2}t}}{\kappa/2} \right)\,, 
\label{eq:an_long}
\end{equation}
which describes the $180^\circ$ splitting of the trajectories sketched in Fig. 1 in the main text. Notice that such $180^\circ$ splitting is also reproduced by Eq.~\eqref{eq:alongchi} at short times $\langle a(t\rightarrow0)\rangle_{\rm long.}=-ig_\parallel\langle \sigma_z\rangle t$, as discussed in the main text. 

{\it Compensation tones -- } It is obvious from Eqs.~\eqref{eq:alpha} and \eqref{eq:beta}, that these results are corrected for finite values of $\bar\chi$ and $\bar g_\parallel$. As it will become apparent in the sections below, such contributions always when considering realistic systems outside of an idealized charge qubit. To get the feeling of how such terms on the dynamics, we can still consider $\chi=0 $ and Eq.~\eqref{eq:an_long} becomes
\begin{equation}
\langle a(t)\rangle_{\rm long.}=-i\left(\bar{g}_\parallel+g_\parallel\langle\tilde\sigma_z(0)\rangle\right)  \frac{1-e^{-\frac{\kappa}{2}t-i\bar\chi}}{i\bar\chi+\kappa/2}\,.
\label{eq:evolution_multilevel}
\end{equation}
This expression shows that the cavity dynamics is controlled by the $\bar g_\parallel$ term which is not sensitive to the value of $\langle\tilde\sigma_z(0)\rangle$. However, inspection Eqs.~\eqref{eq:alpha} and \eqref{eq:beta} indicates that this spurious effects can be compensated choosing by adding an additional driving tone to the resonator tuned to $A_{\rm r}=-2\bar{g}_\parallel$, $\varphi=0$ and $\omega_{\rm d}-\omega_{\rm r}=\bar{\chi}$, as we show in Figs. 3 and 4 in the main text. 

\begin{figure}[!t]
    \centering
    \includegraphics[width=\textwidth]{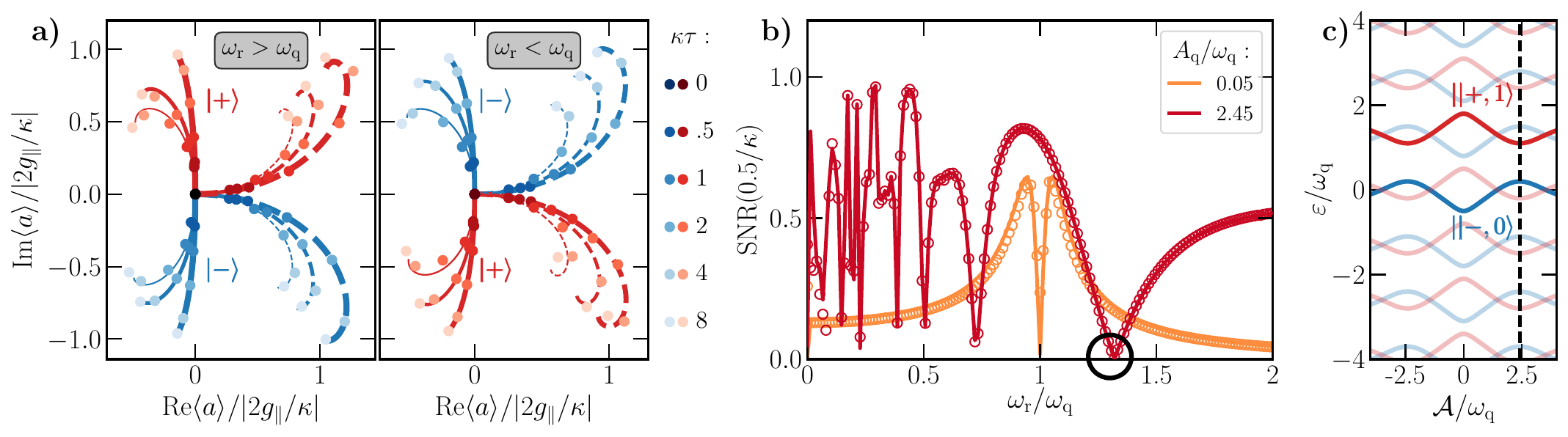}
    \caption{a) Analytical and numerical time trajectories of $\langle a\rangle$, obtained considering $31$ replicas and $15$ levels in the resonator, as in Fig.~2 in the main text. Left panel obtained with $\omega_{\rm r}/\omega_{\rm q}=0.3,0.8,0.9$, right panel is Fig~2a of the main text. We compare the cases $\omega_{\rm r}\gtrless\omega_{\rm q}$, highlighting the inversion of trajectories corresponding to the change of sign of $g_\parallel$ and $\chi$. b) SNR as in Fig. 2. For such small signals, a Fock space of dimension $6$ is sufficient to guarantee convergence of the results. The circle highlights deterioration of the SNR at $\omega_{\rm r}=1.3\omega_{\rm q}$. The analytical curve is obtained with a $g_\parallel$ computed with $81$ replicas. c) Floquet spectrum for $\omega_{\rm r}=1.3\omega_{\rm q}$, obtained with $41$ replicas. The dashed vertical line corresponds to $A_{\rm q}/\omega_{\rm q}=2.45$.}
    \label{fig:detail_longitudinal}
\end{figure}

\subsection{Numerical simulations and comparison to analytical calculations}
The numerical simulations presented in the main text are obtained using the package QuTip~\cite{JOHANSSON_qutip_2013} with the dynamics of the expected value of the operator $\langle a \rangle$ given by Eq.~\eqref{eq:multi_level_H_0}.
The simulations are obtained considering $30$ energy levels in the resonator. The comparison to the analytical calculations of Section~\ref{sec:evolution_a} are done by assuming $a\approx \tilde a$, which is justified by neglecting subleading corrections of order $g_\perp$ in Eq.~\eqref{eq:atildea}. These corrections describe rapid rotation around the main trajectories (not shown).

Concerning the numerical simulation of the ideal charge qubit and the numerical curves shown in Fig. 2 in the main text, we consider  Eq.~\eqref{eq:multi_level_H_0} with $ Q=\varepsilon_0$  and $\mathcal H_{\rm sys}[\varepsilon_0]=\omega_{\rm q}\sigma_z/2+\varepsilon_0\sigma_x$ at the sweet spot $\varepsilon_0=0$. In Fig.~\ref{fig:detail_longitudinal}a we show explicitly that the trajectories associated to $\ket\pm$ switch in the case $\omega_{\rm r}<\omega_{\rm q}$, according to the sign change of $g_\parallel$ and $\chi$ extensively discussed in the main text. In Fig.~\ref{fig:detail_longitudinal}b we compare the SNR at $A_{\rm q}/\omega_{\rm q}=0.05$ (also shown in Fig.~2c in the main text) to the SNR at much larger values than $A_{\rm q}/\omega_{\rm q}=2.45\gg 0.5$. The SNR shows a proliferation of suppression points caused by additional resonances in the Floquet spectrum. We highlight that such supressions are more likely to happen for $\omega_{\rm r}<\omega_{\rm q}$. Figure~\ref{fig:detail_longitudinal}b also shows a suppression arising for $\omega_{\rm r}>\omega_{\rm q}$. However, Figure~\ref{fig:detail_longitudinal}c shows that the corresponding resonance between distant replicas in the Floquet spectrum requires larger values of $A_{\rm q}$. Notice that, for small values of $A_{\rm q}$, the Floquet spectrum in Fig.~\ref{fig:detail_longitudinal} clearly display curvatures and slopes of opposite sign than Fig.~2e in the main text, corresponding to a case where $\omega_{\rm r}<\omega_{\rm q}$. This situation reproduces the heuristics sketched in Fig.~1b in the main text. 

We discuss below the extension to realistic multi-level systems and provide details about the mapping onto Eqs.~\eqref{eq:multi_level_H_0} and~\eqref{eq:effective_H_beyond} for numerical and analytical calculations respectively.

\section{Details about extensions to multi-level systems}
In this concluding section we provide additional details about the extension of our approach to more realistic setups of hybrid and superconducting cQED.

\subsection{Flopping mode}\label{sec:flopping-mode}
In this section we analyze in detail the flopping mode spin-qubit. The Hamiltonian describing a charge trapped in a double quantum dot reads~\cite{Jin2012,frey20212dipole,2012_Petersson,viennot2015coherent,2018_Samkharadze,2018_Mi,2019_Borjans,2020_Harvey-Collard,yu2023strong,2004_Childress,Burkard2006,trif2008spin,cottet2010spin,2012_Hu}
\begin{equation}
\mathcal H_{\rm sys}[\varepsilon_0]=\frac\Delta2\sigma_z+\varepsilon_0\tau_z+t_{\text{sc}} \tau_x - t_{\text{sf}}\, \tau_y \sigma_y\,.
\end{equation}
where the Pauli operators $\sigma_z=\ket{\uparrow}\bra{\uparrow}-\ket{\downarrow}\bra{\downarrow}$, $\tau_z=\ket{L}\bra{L}-\ket{R}\bra{R}$ act on the spin ($\ket{\uparrow,\downarrow}$) and spatial ($\ket{L,R}$) degrees of freedom respectively. A magnetic field induces a Zeeman splitting $\Delta$ between the $\ket{\uparrow,\downarrow}$ spin states, and the amplitudes $t_{\rm sc/sf}$ describe spin-conserving/flipping tunneling between the left ($\ket L$) and right ($\ket R$) dots, which are detuned by the energy $\varepsilon_0$. 

\begin{figure}[b]
    \centering
    \includegraphics[width=.8\textwidth]{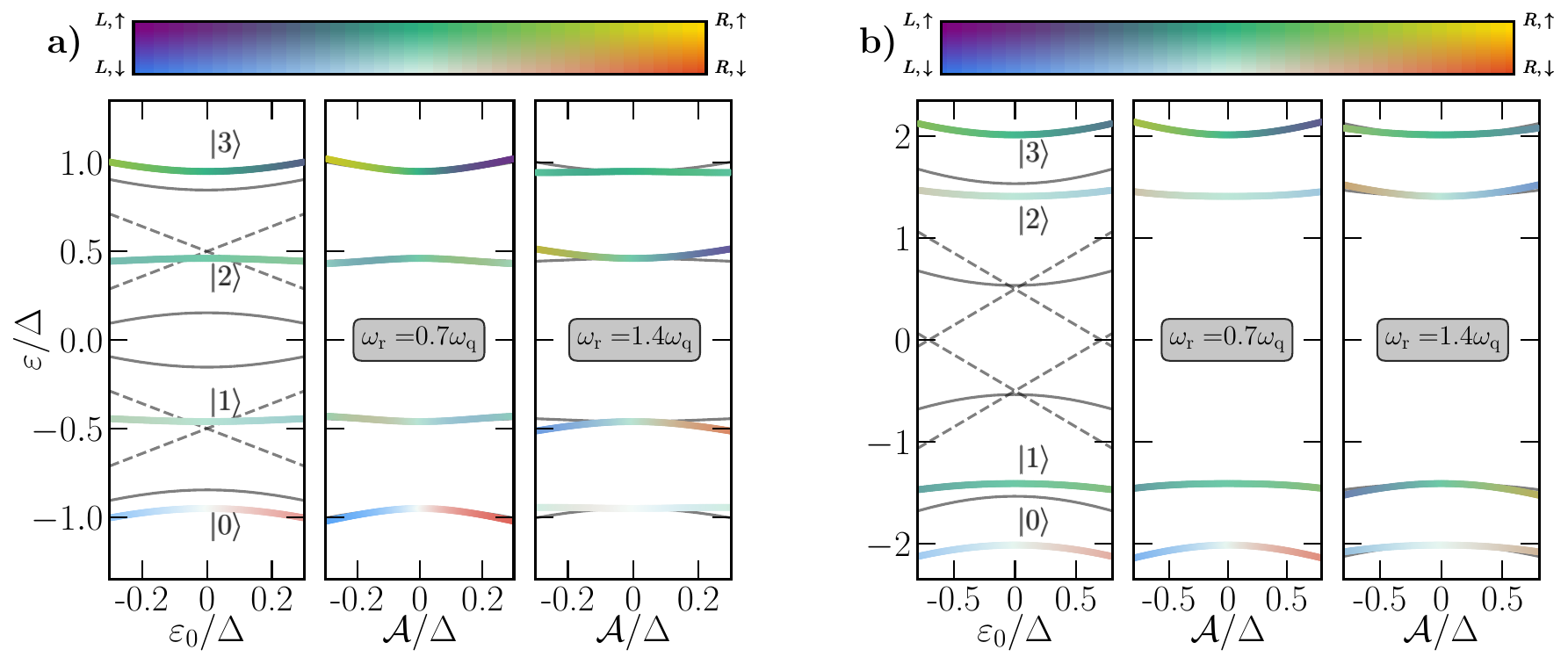}
    \caption{Static and Floquet spectra of the charge-like regime $\Delta>2t_{\rm sc}$ (a) and of the spin-like regime $\Delta<2t_{\rm sc}$ (b). As in the main text, we chose $\mathcal A$ to be real. For the static spectrum (left panel in both (a) and (b)) the grey lines indicate the spectrum in the case $t_{\rm sc}=t_{\rm sf}=0$ (dashed) and $t_{\rm sf}=0$ (solid). In the Floquet spectra grey lines show the limit $\omega_{\rm r}=0$ that corresponds to the static spectrum. Floquet spectra are computed considering $N_{\rm rep}=11$ replicas. Model parameters for charge-like regime (in units of $\Delta$): $t_{\rm sc}=-0.35$, $t_{\rm sf}=0.43$. The spin-like regime is the same as the one shown in the main text. }
    \label{fig:flopping_case}
\end{figure}

In Fig.~\ref{fig:flopping_case} we analyze the different operational regimes of this qubit, where the static spectrum is obtained by varying the detuning $\varepsilon_0$ and it is compared to the Floquet spectrum when we consider Eq.~\eqref{eq:multi_level_H_0} and drive the detuning ($Q=\varepsilon_0$) at the sweet spot $\varepsilon_0=0$. The charge-like regime is obtained for $\Delta>2t_{\rm sc}$ (Fig.~\ref{fig:flopping_case}a) and it is compared to the spin-like regime ($\Delta>2t_{\rm sc}$, Fig.~\ref{fig:flopping_case}b), which is discussed in the main text. 

In the charge-like case, the magnetic field  is so strong that the spin sectors are well separated and the two lower levels have mostly a $\ket{\downarrow}$ character while the two upper ones mostly a $\ket{\uparrow}$ character. Both the two upper and lower branches of the spectrum reproduce all the qualitative features of the ideal charge qubit. For instance, the curvature of the Floquet spectrum  change sign when $\omega_{\rm q} \simeq \omega_{\rm r}$. 

In the spin-like regime $\Delta<2t_{\rm sc}$ the magnetic field slightly separates the $\ket{\downarrow}$ and $\ket{\uparrow}$ states while the spin-conserving tunneling $t_{\rm sc}$ splits the antisymmetric state $\ket{-}=(\ket L-\ket R)/2$ from the symmetric state $\ket{+}=(\ket L+\ket R)/2$, see Fig.~\ref{fig:flopping_case}b. At zero detuning ($\varepsilon_0=0$) we would obtain, without spin-orbit coupling, that the transition between the two lower energy levels $\ket{-,\downarrow}$ and $\ket{-,\uparrow}$ would consist in a pure spin flip. However, the presence of spin-orbit coupling ($t_{\rm sf}\neq0$) allows to couple charge and spin. As a consequence, driving the detuning has a different impact on the curvatures of the $\ket 0$ and $\ket 1$ states. In particular, switching between the conditions $\omega_{\rm r}\gtrless\omega_{\rm q}$ also causes, following the definitions of the effective couplings Eq.~\eqref{eq:gparallel_chi_multilevel}, a change of sign of the $g_\parallel$  (not shown) even though $\bar g_\parallel$ stays negative.  In the main text, we show in Fig.~3b how the compensation procedure described in Section~\ref{sec:compensation} allows to isolate the longitudinal dynamics.

For all analytical and numerical applications, the quasi-energies and the Floquet modes are computed using $\mathcal H_{\mathcal F}$~\eqref{eq:projected_floquet_H_multi_level} with $Q=\tau_z$ and including up to $N_{\rm rep}=31$, whereas $g_\parallel,\bar{g}_\parallel,\chi,\bar{\chi}$ are obtained through Eq.~\eqref{eq:gparallel_chi_multilevel}. 

\subsection{Transmon}
\label{sec:transmon}
In this section we give the details about the model used to describe longitudinal and dispersive readout in the tunable transmon, framing the study of Ref.~\cite{ikonen2019qubit} in the Floquet approach presented in our work. We consider always the general  situation described by Eq.~\eqref{eq:multi_level_H_0}, in which the system Hamiltonian of the tunable transmon is~\cite{koch2007charge-insensitive, ikonen2019qubit}
\begin{equation}
\mathcal{H}_{\rm sys}[Q]=4 E_C \left( n+\frac{Q}{8E_C}\right)^2 - E_J \cos \phi\,,
\end{equation}
where we have introduced $Q=-8E_Cn_g$ to find the usual form of the coupling (see Eq.~\eqref{eq:coupling_transmon}). The charge $n$ and phase $\phi$ operators are conjugated variables ($[n,\phi]=i$) and $E_J$ is the effective Josephson energy. The parameters of the transmon are adapted from Ref.~\cite{ikonen2019qubit} and are shown in Table~\ref{tab:table_transmon_fluxonium}. 

\begin{figure}[!t]
    \centering
    \includegraphics[width=\textwidth]{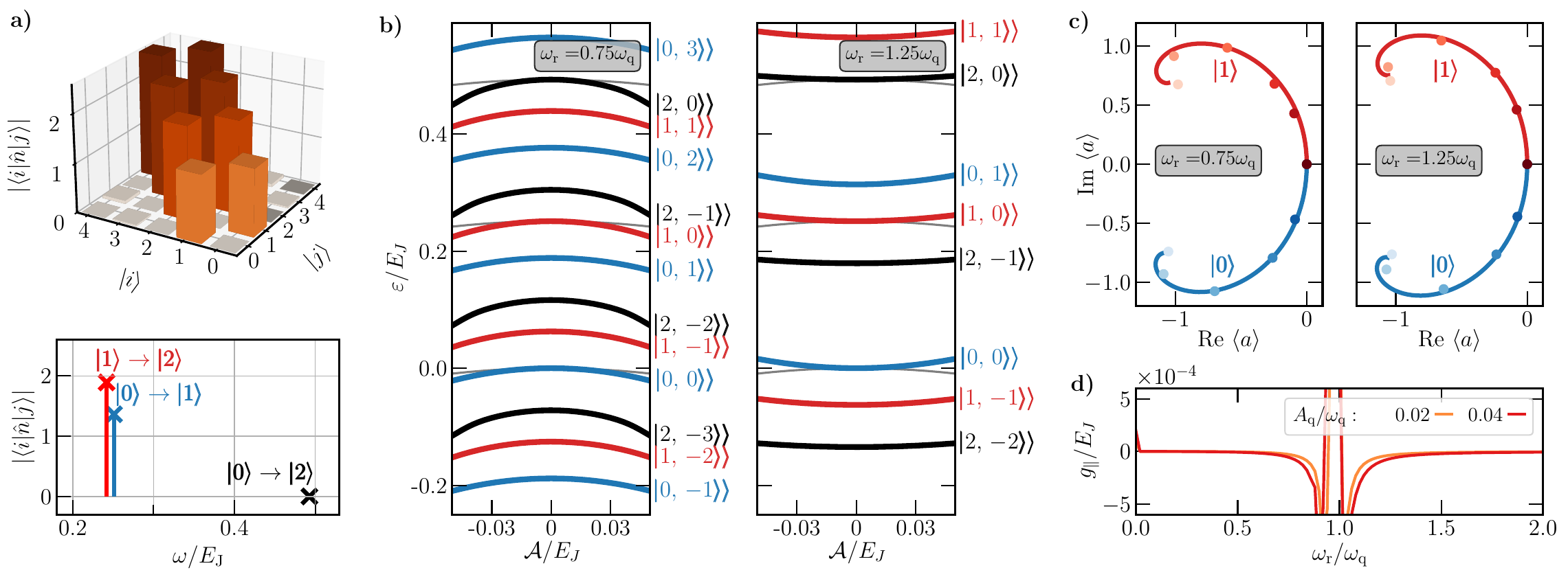}
    \caption{a) Matrix elements of the charge operator $n$ for the transmon. b) Floquet spectrum for $\omega_{\rm r} \gtrless \omega_{\rm q}$. Grey lines represent the spectrum in the limit $\omega_{\rm r} \rightarrow 0$. The Floquet spectrum is computed with $N_{\rm rep}=31$ and considering $20$ states in the transmon to ensure convergence in function of the number of states and replicas. c) Evolution of the cavity pointer states as in Fig. 4 of the main text. d) $g_\parallel$ in function of the cavity frequency for different value of $A_{\rm q}$.}
    \label{fig:floquet_spectrum_transmon}
\end{figure}

To obtain the quasi-energies for this system we write $\mathcal H_{\mathcal F}$~\eqref{eq:projected_floquet_H_multi_level} in the basis of the transmon eigenstates $\{ \ket{j} \}$, with the transmon energies $E_j$ and the matrix elements $n_{ij}=\braket{i \vert n \vert j}$ (see Fig.~\ref{fig:floquet_spectrum_transmon}a) obtained using the package Scqubits~\cite{scqubits1,scqubits2} with the parameters listed in Table~\ref{tab:table_transmon_fluxonium}. We then consider just the first $20$ levels of the transmon to build $n=\sum_{i,j=0,\ldots,2}n_{ij}\ket{i}\bra{j}$ and we truncate $\mathcal H_{\mathcal F}$ up to $N_{\rm rep}=31$ replicas. Solving the spectrum of $\mathcal H_{\mathcal F}$, we compute the Floquet modes and their quasi-energies, which are plotted in   Fig.~\ref{fig:floquet_spectrum_transmon}b. Projecting into the $\{ \ket{0},\ket{1} \}$ subspace, we obtain $g_\parallel,\,\bar{g}_\parallel,\,\chi,\,\bar{\chi}$ through Eq.~\eqref{eq:gparallel_chi_multilevel}. 

Differently from the spin-qubit discussed in Section~\ref{sec:flopping-mode}, the weak anharmonicity of the transmon makes the longitudinal coupling always negative far from $\omega_{\rm q}=\omega_{\rm d}$. To show this explicitely, we consider the asymptotic expressions, valid in the regime $E_J/E_C\rightarrow \infty$~\cite{koch2007charge-insensitive}
\begin{equation}
E_j \simeq -E_J+\sqrt{8E_JE_C} \left( j+\frac12 \right)-\frac{E_C}{12}(6j^2+6j+3)\,.
\end{equation}
and
\begin{equation}
\begin{aligned}
&  \braket{j+1 \vert n \vert j} \approx i \sqrt{\frac{j+1}{2}}\left( \frac{E_J}{8E_C}\right)^{1/4}\,, \\
&\braket{j+k \vert n \vert j} \approx 0\,,
\end{aligned}
\end{equation}
with $\vert k \vert > 1$. Inserting $\left. \frac{\partial \mathcal H_{\rm sys}}{\partial Q} \right\vert_{n_g=0}=n$ into Eq. ~\eqref{eq:multi_level_H_0}, we obtain
\begin{equation}
\mathcal{H}=\sum_j E_j \ket{j}\bra{j}+g_\perp (a+a^\dagger) n + A_{\rm q}\cos(\omega_{\rm d}t)n +\omega_{\rm r}a^\dagger a+A_{\rm r}\cos(\omega_{\rm d}t+\varphi)(a+a^\dagger)\,.
\label{eq:coupling_transmon}
\end{equation}
In this case $\mathcal H_{\mathcal F}$ reads
\begin{equation}
\mathcal H_{\mathcal F}=
\begin{pmatrix}\ldots\\
 & E_{2}+\omega_{{\rm d}} & 0 & 0 & 0 & \frac{\mathcal{A^{*}}}{2}n_{12}^{*} & 0 &  0 & 0 & 0\\
 & 0 & E_{1}+\omega_{{\rm d}} & 0 & \frac{\mathcal{A^{*}}}{2}n_{12} & 0 & \frac{\mathcal{A^{*}}}{2}n_{01}^{*} & 0 &  0 & 0\\
 & 0 & 0 & E_{0}+\omega_{{\rm d}} & 0 &  \frac{\mathcal{A^{*}}}{2}n_{01} & 0 & 0 & 0 &  0\\
 & 0 &  \frac{\mathcal{A}}{2}n_{12}^{*} & 0 & E_{2} & 0 & 0 & 0 &  \frac{\mathcal{A^{*}}}{2}n_{12}^{*} & 0\\
 &  \frac{\mathcal{A}}{2}n_{12} & 0 &  \frac{\mathcal{A}}{2}n_{01}^{*} & 0 & E_{1} & 0 &  \frac{\mathcal{A^{*}}}{2}n_{12} & 0 &  \frac{\mathcal{A^{*}}}{2}n_{01}^{*}\\
 & 0 &  \frac{\mathcal{A}}{2}n_{01} & 0 & 0 & 0 & E_{0} & 0 &  \frac{\mathcal{A^{*}}}{2}n_{01} & 0\\
 &  0 & 0 & 0 & 0 &  \frac{\mathcal{A}}{2}n_{12}^{*} & 0 & E_{2}-\omega_{{\rm d}} & 0 & 0\\
 & 0 &  0 & 0 &  \frac{\mathcal{A}}{2}n_{12} & 0 &  \frac{\mathcal{A}}{2}n_{01}^{*} & 0 & E_{1}-\omega_{{\rm d}} & 0\\
 & 0 & 0 &  0 & 0 &  \frac{\mathcal{A}}{2}n_{01} & 0 & 0 & 0 & E_{0}-\omega_{{\rm d}}\\
 &  &  &  &  &  &  &  &  &  & \ldots
\end{pmatrix}\,, \end{equation}
and second order perturbation theory in $\mathcal A$ gives the quasi-energies
\begin{equation}
\begin{aligned}
\varepsilon_0 &=E_0-\vert n_{01} \vert^2\frac{\vert \mathcal A \vert^2}{2}\left(\frac{1}{\omega_{\rm q}-\omega_{\rm d}}+\frac{1}{\omega_{\rm q}+\omega_{\rm d}}\right)\,,\\
\varepsilon_1 &=E_1+\vert n_{01}\vert^2 \frac{\vert \mathcal A \vert^2}{2} \left( \frac{1}{\omega_{\rm q}-\omega_{\rm d}}+\frac{1}{\omega_{\rm q}+\omega_{\rm d}}-2\frac{1}{\omega_{\rm q}-E_C-\omega_{\rm d}}-2\frac{1}{\omega_{\rm q}-E_C+\omega_{\rm d}}\right)\,,\\
\end{aligned}
\end{equation}
where we have used that $\vert n_{12} \vert^2=2\vert n_{01} \vert^2$. From these expressions, we can compute $g_\parallel$ and $\chi$ through Eq.~\eqref{eq:gparallel_chi_multilevel}
\begin{align}\label{eq:transmon_g} 
g_\parallel^{(0)} &= g_\perp A_{\rm q}\vert n_{01}\vert^2\frac12\left(\frac{1}{\omega_{\rm q}-\omega_{\rm d}}+\frac{1}{\omega_{\rm q}+\omega_{\rm d}}-\frac{1}{(\omega_{\rm q}-E_C)-\omega_{\rm d}}-\frac{1}{(\omega_{\rm q}-E_C)+\omega_{\rm d}}\right)\,,&
\chi^{(0)}&=\frac{2g_\perp}{g_\parallel^{(0)}}A_{\rm q} 
\,,
\end{align}
with $\omega_{\rm q}=E_1-E_0$. We remark that, neglecting the counter-rotating terms $\propto (\omega_{\rm q}+\omega_{\rm d})^{-1}$, $(\omega_{\rm q}-E_C+\omega_{\rm d})^{-1}$, Eq.~\eqref{eq:transmon_g} reproduces the well known expression for the dispersive coupling~\cite{blais_cQED_2021}
\begin{equation}
\chi^{(0)} \approx -g_\perp^2 \vert n_{01} \vert^2 \frac{E_C}{(\omega_{\rm q}-\omega_{\rm d})(\omega_{\rm q}-\omega_{\rm d}-E_C)}\,.
\end{equation}
This expression shows that both the longitudinal and the dispersive coupling, for $E_C < \vert \omega_{\rm q} - \omega_{\rm d} \vert $, are negative. 

\subsection{Fluxonium}
\label{sec:fluxonium}
\begin{figure}[t]
    \centering
    \includegraphics[width=\textwidth]{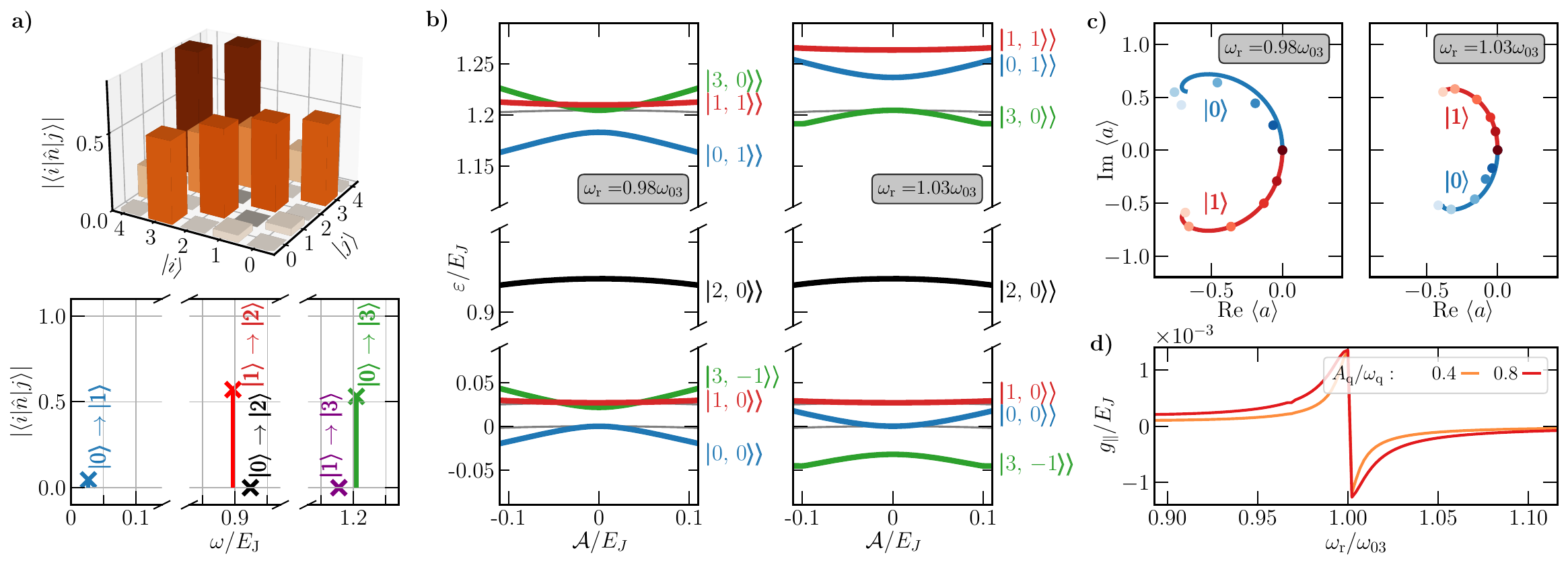}
    \caption{a) Matrix elements of the charge operator $n$ for the fluxonium. b) Floquet spectrum for $\omega_{\rm r} \gtrless \omega_{03}$. Grey lines represent the spectrum in the limit $\omega_{\rm r} \rightarrow 0$. The Floquet spectrum is computed with $N_{\rm rep}=31$ and considering $20$ states in the fluxonium to ensure convergence in function of the number of states and replicas. c) Evolution of the cavity pointer states as in Fig. 4 of the main text. d) $g_\parallel$ in function of the cavity frequency for different value of $A_{\rm q}$.}
    \label{fig:floquet_spectrum_fluxonium}
\end{figure}

In this section we give the details about the model used to describe longitudinal and dispersive readout in the fluxonium.
The system Hamiltonian for the fluxonium~\cite{manucharyan2009fluxonium} reads
\begin{equation}
\mathcal{H}_{\rm sys}[Q]=4E_C \left(n+\frac{Q}{8E_C} \right)^2 - E_{J} \cos \left( \phi - 2 \pi \frac{\Phi_{\rm ext}}{\Phi_0} \right) + \frac{1}{2}E_L \phi^2\,.
\end{equation}
As for the transmon, we use the package Scqubits~\cite{scqubits1,scqubits2}, with the parameters of Table~\ref{tab:table_transmon_fluxonium}, to find the uncoupled eigenstates $\{ \ket{j} \}$ and the matrix elements $n_{ij}$. As for the transmon, we consider only the first $20$ levels of the fluxonium to build the charge operator in the Floquet Hamiltonian and we truncate $\mathcal H_{\mathcal F}$ up to $N_{\rm rep}=31$ to obtain the quasi-energies and the Floquet modes.  Considering the $\{ \ket{0},\ket{1} \}$ subspace, we obtain $g_\parallel,\,\bar{g}_\parallel,\,\chi,\,\bar{\chi}$ through Eq.~\eqref{eq:gparallel_chi_multilevel}. 
As shown in Fig.~\ref{fig:floquet_spectrum_fluxonium}b-d, since at leading order the ground state couples only with $\ket{3}$, we observe a change of sign in the $g_\parallel$ at $\omega_{\rm d} \simeq \omega_{03}$.
\begin{table}[!hbt]
    \centering
    \begin{tabular}{c|cc}
         & Transmon & Fluxonium \\
         \hline
        $E_C/E_J$ & $8.4 \cdot 10^{-3}$ & 0.2 \\
        $E_L/E_J$ & 0 & 0.1 \\
        $\Phi_{\rm ext}/\Phi_0$ & 0 & 0.5 \\ 
        $g_\perp/E_J$ & $3 \cdot 10^{-3}$ & 0.01 \\
        $\kappa/E_J$ & $6.4\cdot 10^{-5}$ & $1 \cdot 10^{-3}$ \\
        $A_{\rm q}/\omega_{\rm q}$ & 0.04 & 0.8 \\
        $\omega_{\rm q}/E_J$ & 0.25 & 0.03 \\
        $\omega_{\rm r}/\omega_{\rm q}$ & 0.75 & 46\\
    \end{tabular}
    \caption{Parameters used to obtain Fig. 4 in the main text.}
    \label{tab:table_transmon_fluxonium}
\end{table}

\end{document}